\documentclass[preprint]{article}
\newcommand{\be}{\begin{equation}} \newcommand{\ee}{\end{equation}} 
\newcommand{\bea}{\begin{eqnarray}}\newcommand{\eea}{\end{eqnarray}}

\begin{document}
\title{ Supersymmetric Many-particle Quantum Systems with Inverse-square
Interactions}
\author{ Pijush K. Ghosh\footnote{{\bf email:}
pijushkanti.ghosh@visva-bharati.ac.in}}
\date{Department of Physics, Siksha-Bhavana,\\ 
Visva-Bharati University,\\
Santiniketan, PIN 731 235, India.}
\maketitle
\begin{abstract} 
The development in the study of supersymmetric many-particle quantum
systems with inverse-square interactions is reviewed. The main emphasis
is on quantum systems with dynamical $OSp(2|2)$ supersymmetry. Several
results related to exactly solved supersymmetric rational Calogero model,
including shape invariance, equivalence to a system of free superoscillators
and non-uniqueness in the construction of the Hamiltonian, are presented in
some detail. This review also includes a formulation of pseudo-hermitian
supersymmetric quantum systems with a special emphasis on rational Calogero
model. There are quite a few number of many-particle quantum systems with
inverse-square interactions which are not exactly solved for a complete set
of states in spite of the construction of infinitely many exact eigen
functions and eigenvalues.
The Calogero-Marchioro model with dynamical $SU(1,1|2)$ supersymmetry and
a quantum system related to short-range Dyson model belong to this class
and certain aspects of these models are reviewed. Several other related and
important developments are briefly summarized.
\end{abstract}

\tableofcontents

\vspace{0.2in}
\section{Introduction}

Supersymmetry plays an important role in understanding unification of
different fundamental forces in nature. In the earlier days of the 
development of the subject, supersymmetric quantum mechanics was
introduced to study different aspects of supersymmetric quantum field
theory within a much simpler and tractable framework\cite{witten, freedman}.
However, the importance of developing supersymmetric quantum mechanics as
an independent subject was soon apparent to physicists as well as
mathematicians working in this area. Several aspects of supersymmetric
quantum mechanics have been investigated over the last three decades
and many of these important developments are summarized in a few review
articles and books, for example, in Refs. \cite{cooper,bagchi,asim}. 

A study on supersymmetric quantum systems with many degrees of freedom was
initiated in Ref. \cite{rit}, just within two years of the development of
the subject. However, unlike systems with one bosonic and one fermionic
degrees of freedom, not many exactly solved supersymmetric many-particle
quantum systems are known. This is because of the following reasons.
Supersymmetric systems with one bosonic and one fermionic degrees of freedom
are described in terms of two Hamiltonians, known as supersymmetric
partner Hamiltonians. The eigen functions and eigenvalues of one of these
Hamiltonians can be obtained from the other one and the vice verse, by using
the intertwining relations arsing from the underlying superalgebra\cite{cooper}.
The complete eigen spectra can be obtained algebraically provided the
partner Hamiltonians satisfy the condition of shape invariance and examples
of such quantum systems are abundant in the literature\cite{cooper,bagchi,asim}.

In the case of systems with $N$ bosonic and $N$ fermionic degrees of freedom,
the supersymmetric Hamiltonian can be expressed in the fermionic basis as a
$2^N \times 2^N$ block-diagonal matrix with $N+1$ components
corresponding to $N+1$ fermionic sectors on the diagonal\cite{andrianov}.
Each sector is characterized by the total fermion number.
The zero-fermion and the $N$-fermion sectors are described by two different
Hamiltonians, while matrix-operators of different dimensions appear in rest
of the $N-1$ sectors\cite{andrianov}. The underlying super algebra allows
an one-to-one correspondence between the eigenfunctions and the eigenvalues
of these two Hamiltonians for only systems with one bosonic and
one fermionic degrees of freedom, i.e. $N = 1$. It is true that part of the
spectrum of a block-operator corresponding to a fixed fermionic sector
coincide with the spectra of neighbouring blocks for systems with $N >1$.
However, a successful implementation of the condition of shape invariance,
which is essential for showing exact solvability, meets with difficulty for
arbitrary $N$. In general, the diagonalization of the block-diagonal matrix
operator is also a highly non-trivial task. These technical limitations
appear as the main stumbling block for finding physically relevant exactly
solved many-particle supersymmetric quantum systems. Nevertheless, a class of
exactly solved many-particle supersymmetric quantum systems exists for which
the many-body interaction varies inverse-squarely.

Inverse-square interactions appear naturally in physical systems in two or
higher space dimensions as the centrifugal barrier. There are several low
dimensional condensed matter systems governed by inverse-square
interactions\cite{kuramoto}. Systems with purely inverse-square interaction
are also example of conformal mechanics\cite{dff,fr} which are relevant in
diverse branches of physics.
The Calogero model\cite{cs} is an exactly solved system of $N$ particles
interacting on a line through pair-wise inverse-square and harmonic 
interactions. The model is exactly solved even if the pair-wise harmonic
interaction is replaced by a common confining harmonic potential or a periodic
version of the inverse-square interaction without the confining harmonic term is
considered\cite{cs1}. Several other generalizations of the original Calogero
model have been considered over a period of more than four decades and this
class of exactly solved many-particle systems is generically known as
Calogero-Moser-Sutherland systems. There are excellent reviews on the topic,
for example, in Refs. \cite{pr,poly,revpoly,pasquier,zagreb}. The study of
Calogero-Moser-Sutherland systems have produced many interesting results
which are relevant in the context of a diverse branches of physics,
including exclusion statistics\cite{murthy,ha}, quantum Hall effect\cite{qhe},
Tomonaga-Luttinger liquid\cite{tll}, quantum chaos\cite{qc}, electric
transport in mesoscopic systems\cite{meso}, novel correlations\cite{date},
spin-chains\cite{hs,hs1,mina,ahn,biru1,biru2,biru3,biru4,spain2} etc. These
developments are also important in the  context of mathematical physics, for
example,
algebraic and integrable structure\cite{wadati,awata,sw1}, mapping of rational
model to Calogero model with Coulomb-like potential\cite{calo_coul},
self-adjoint extensions\cite{bbm,tsutsui}, equivalence to a system of free
oscillators\cite{pani,poland}, collective field formulation of many-particle
systems\cite{joshua} etc.

A supersymmetric version of the rational Calogero Model was introduced by
Freedman and Mende\cite{fm} in 1990. Various aspects of this class of
supermodels have been studied over the last two decades, enriching a general
understanding of the integrable structure of many-particle supersymmetric
quantum systems\cite{fm,bh,turb,susy,nonu,ss,me,sasa,rus,ioffe2,gala2}.
Apart from being examples of exactly solved quantum systems, the importance
of these models is due to their relevance in the study of black-holes\cite{gib},
Seiberg-Witten theory\cite{sw}, matrix models and string
theory\cite{atish,gates,verlinde,sameer,agarwal}, collective field
theory\cite{tonder,rodrigues,jevicki}, many-particle superconformal quantum
mechanics\cite{pkg,nw,sc,belu,wen,ohta,olaf1,olaf2,olaf3,gala1,gala3,gala4,gala5,kiri,review_scm},
superpolynomials\cite{sp0,sp1,sp2,sp3,sp4}, pseudo-hermitian supersymmetric
Calogero models\cite{piju_latest} etc.

The main emphasis of this topical review article is on systems with $OSp(2|2)$
dynamical supersymmetry. The rational $A_{N+1}$ and $BC_{N+1}$ Calogero models
belong to this class and an algebraic construction of spectra of these two
models is discussed. The eigen value problem of the Hamiltonians
appearing in the zero and $N$ fermion sectors of the supersymmetric Hamiltonian
is solved by using the Dunkl operators and the idea of shape
invariance\cite{me}. The supersymmetric Hamiltonian can also be mapped to a
system of free superoscillators through a non-unitary similarity
transformation, facilitating a construction of the complete set of states
from the free superoscillator basis\cite{susy}. However, only those free
superoscillator states are acceptable which are invariant under the discrete
symmetries of the original many-body Hamiltonian. The discrete symmetries of
the rational $A_{N+1}$ and $BC_{N+1}$-type Calogero models are different.
Thus, although both of these Hamiltonians can be mapped to the same free
superoscillator Hamiltonian, the eigen spectra are not identical. 

The mathematical aspects of $OSp(2|2)$ supersymmetry is described in Refs.
\cite{superlie,rev,dhowker,scasi}. It is known that $OSp(2|2)$ admits
`typical' as well as `atypical' representations. The quadratic and the
cubic Casimir operators are necessarily zero in the `atypical' representation.
The supersymmetric Calogero model introduced by Freedman and Mende\cite{fm}
corresponds to `typical' representation of $OSp(2|2)$ group. A supersymmetric
version of the rational Calogero model corresponding to the `atypical'
representation of the $OSp(2|2)$ group is presented in this article\cite{nonu}.

The study of ${\cal{PT}}$ symmetric non-hermitian quantum system has
received considerable attention in the literature over the last
decade\cite{bend,ali,quasi,ddt}. Several pseudo-hermitian quantum systems
with an exact description of the norm in the Hilbert space have been
considered\cite{piju2,me1,piju4}. The rational Calogero model and its
variants have also been considered in the literature within the same
context\cite{ptcsm,oldpiju,europe,smith}. A general construction of
pseudo-hermitian supersymmetric quantum systems and rational Calogero
model is presented\cite{piju_latest}. 

There are many higher dimensional generalizations of the rational Calogero
model\cite{marchio,mbs,mri}.  Although infinitely many exact eigenstates can
be obtained analytically for these models, not a single Hamiltonian belonging
to this class is known to be exactly solved for a complete set of states.
The Calogero-Marchioro model\cite{marchio} is one such example of a `partially
solved' system, which has interesting connections with complex random matrix
theory\cite{kray,rsu,go}, two dimensional Bose systems\cite{rsu}, quantum Hall
effect\cite{kklz}, quantum dot\cite{imsc}, collective field
theory\cite{joshua1}. The construction of supersymmetric
Calogero-Marchioro model in arbitrary $D$ space dimensions with $OSp(2|2)$
symmetry is presented. It is also shown that the same model in two space
dimensions has an extended $SU(1,1|2)$ superconformal symmetry for arbitrary
number of particles and the generic value of the coupling constant\cite{pkg}.   

There are generalizations of rational Calogero model where each particle
interacts only with the nearest-neighbour and the next-nearest-neighbour
particles through an inverse-square interaction\cite{nni,1nni}. Infinitely
many exact eigenstates can be obtained analytically for this system for
open or periodic boundary condition. However, these states do not form a
complete set of states. This model has a close connection with random banded
matrix theory describing short-range Dyson model\cite{rbm} and spin
chains\cite{degu,spain1,spain3,spain4}. A supersymmetric
version of this model with dynamical $OSp(2|2)$ symmetry is presented in this
article\cite{degu}. The supersymmetric Hamiltonian can also be equivalently
described as an interacting system of $N$ particles with spin degrees of
freedom, where each particle interacts with only its nearest-neighbour and
the next-nearest-neighbour through inverse-square interaction. Such an
interpretation
is admissible by expressing fermionic operators in terms of Pauli matrices
via Jordan-Wigner transformation. In an appropriate limit, the spin degrees
of freedom can be completely decoupled from the bosonic degrees of freedom
and models of nearest-neighbour $XY$ model in an external magnetic field
on a non-uniform lattice can be obtained.

The plan of presenting the results in this review are the following. The basic
formalism for discussing many-particle supersymmetric quantum systems is
discussed in the next section. The supersymmetric rational $A_{N+1}$ Calogero
model is studied in section 3. The eigen spectra of the Hamiltonians
corresponding to the zero and the $N$ fermion sectors are obtained with the
help of Dunkl operators and shape invariance. The section 4 is an
exploration of systems with $OSp(2|2)$ symmetry. The non-uniqueness in
constructing model Hamiltonians with $OSp(2|2)$ symmetry is pointed out.
It is also shown that the bosonic $O(2,1) \times U(1)$ sub-algebra of
$OSp(2|2)$ can be exploited to show an equivalence between the many-particle
Hamiltonian and a system of free superoscillator. In section 5, a general
construction of pseudo-hermitian supersymmetric quantum system with rational
Calogero model as an example is given. The supersymmetric version of the
Calogero-Marchioro models is studied in section 6, while models related to
short-range Dyson model is presented in section 7. Finally, topics not
included in this review are briefly summarized in section 8. This section
also includes discussions on open problems and a summary of the results
obtained. Several appendices containing mainly mathematical prerequisites
(except section 9.4) are included in section 9. The results for supersymmetric
rational $BC_{N+1}$ Calogero model is included in section 9.4.

A few points regarding the unit and convention used in this article.
The velocity of light, the Planck's constant and mass of identical particles
are taken to be unity. The angular frequency of the common harmonic confining
term is denoted as $\omega$, which is assumed to be equal to unity, unless
mentioned otherwise explicitly. All the supersymmetric Hamiltonians considered
in this review article correspond to `typical' representation of $OSp(2|2)$,
except for Hamiltonians in section 4.1. The supersymmetric phase is
characterized by conditions on the parameters arising from the criteria that
single particle momentum operators are  self-adjoint for the normalizable
zero-energy groundstate wavefunction.

\section{Supersymmetry: General Formalism}

The super-algebra that is relevant in the description of a supersymmetric
quantum system has the general form\cite{witten},
\be
\{Q_{\alpha}, Q_{\beta} \} = 2 \delta_{\alpha \beta} H, \ \
\left [ H, Q_{\alpha} \right ]= 0, \alpha, \beta= 1, 2, \dots {\cal{N}},
\label{salgebra}
\ee
\noindent where $Q_{\alpha}$ are ${\cal{N}}$ real supercharges and $H$ is
the supersymmetric Hamiltonian. The super-algebra in Eq. (\ref{salgebra})
can be shown to be a sub-algebra of the relativistic two dimensional ${\cal{N}}$
extended super-algebra\cite{rit}. Higher symmetry may persists for specific
quantum systems. The main focus in this review article is on $OSp(2|2)$
supersymmetry with ${\cal{N}}=2$ and on $SU(1,1|2)$ supersymmetry with
${\cal{N}}=4$. The relevant superalgebra are given in appendix-II and
Appendix-III, respectively.

The supersymmetric Hamiltonian $H$ depends on the bosonic as well as fermionic
co-ordinates. The positions and the momenta operators correspond
to the bosonic operators, while fermionic co-ordinates generally describe
internal degrees of freedom which may be identified as spin for some specific
cases. A minimal realization of the superalgebra (\ref{salgebra}) for
${\cal{N}}=2$ describes a single particle suspersymmetric quantum
system, where the fermionic degrees of freedom are realized in terms
of the Pauli matrices. Elements of the Clifford algebra are used to realize
the superalgebra for a many-particle quantum system. A general discussion in
this regard on Clifford algebra and realization of ${\cal{N}}=2$ super-algebra
are discussed in the next two sections.

\subsection{Clifford Algebra}

The real Clifford algebra of $2 N$ entities $\xi_p$ is described by the
relations\cite{clifford},
\be
\{\xi_p, \xi_q \} = 2 \delta_{pq}, \ \ p, q=1, 2, \dots 2N. 
\label{cliff}
\ee
\noindent An idempotent operator $\xi_{2N+1}$ may be introduced in terms
of these elements as,
\be
\xi_{2N+1} =  \left ( -i \right )^N \ \xi_1 \xi_2 \dots \xi_{2 N-1} \xi_{2 N},
\label{5gamma}
\ee
\noindent which anti-commutes with all the $\xi_p$'s. The operator $\xi_{2N+1}$
facilitates the construction of the projection operators
$\xi_{2N+1}^{\pm}$,
\bea
&& \xi_{2N+1}^{\pm}= \frac{1}{2} ( 1 \pm \xi_{2N+1} ), \nonumber \\
&& (\xi_{2N+1}^{\pm})^2 = \xi_{2N+1}^{\pm},
\ \ \xi_{2N+1}^{\pm} \xi_{2N+1}^{\mp}=0.
\eea
\noindent A particular realization of the generators of the $O(2N)$ group
is in terms of the elements of the Clifford algebra,
\be
J_{pq} = \frac{i}{4} \left [\xi_p, \xi_q \right ].
\label{generator}
\ee
\noindent The generators of the group $O(2N+1)$ are realized by including
$\xi_{2N+1}$ and allowing $p, q=1, 2, \dots 2N+1$ in Eq. (\ref{generator}).
A matrix representation of the elements $\xi_p$ is given in Appendix-I.

A set of fermionic variables $\psi_i$ and their conjugates $\psi_i^{\dagger}$
may be introduced in terms of $\xi_i$'s as,
\be
\psi_i = \frac{1}{2} \left ( \xi_i - i \xi_{N+i} \right ), \ \
\psi_i^{\dagger} = \frac{1}{2} \left ( \xi_i + i \xi_{N+i} \right ),\ \
i, j=1, 2,\dots N,
\ee
\noindent which satisfy the complex Clifford algebra,
\be
\{\psi_i, \psi_j \} = \{\psi_i^{\dagger}, \psi_j^{\dagger} \} =0, \ \
\{\psi_i, \psi_j^{\dagger} \} = \delta_{ij}.
\ee
\noindent The fermionic permutation operator is defined\cite{bh} as,
\be
K_{ij} := \frac{1}{2} \left [ \psi_i^{\dagger} - \psi_j^{\dagger},
\psi_i - \psi_j \right ] = 1 -
\left (\psi_i - \psi_j \right ) \left (\psi_i^{\dagger} -
\psi_j^{\dagger} \right ),
\label{fermi_ex}
\ee
\noindent which acts on fermionic operators and satisfies the
following properties:
\be
K_{ij} \psi_i^{\dagger} = \psi_j^{\dagger} K_{ij}, \ \ K_{ij} \psi_k^{\dagger}
= \psi_k^{\dagger} K_{ij} \ for \ k \neq i, j, \ \ K_{ij}^2 = 1.
\label{kal}
\ee
\noindent The first two relations in Eq. (\ref{kal}) may be re-written in a
compact form as\cite{rus},
\bea
&& K_{ij} \psi_k^{\dagger} = \sum_{l} T_{(ij)lk} \psi_l^{\dagger}
K_{ij}\nonumber \\
&& T_{(ij)lk} \equiv \delta_{lk} - \delta_{li} \delta_{ki} -
\delta_{lj} \delta_{kj} + \delta_{li} \delta_{kj} + \delta_{lj} \delta_{ki}.
\eea
\noindent The action of $K_{ij}$ on any bosonic operator or variable leaves
it unchanged.

The fermionic vacuum $|0 \rangle$ and its conjugate $|\bar{0}
\rangle$ in the $2^N$ dimensional fermionic Fock space are
defined as, $\psi_i |0 \rangle = 0, \psi_i^{\dagger} |\bar{0} \rangle=0 \
\forall \ i$. The fermion number operator $n_i=\psi_i^{\dagger} \psi_i$
corresponding to the $i^{th}$ fermion has the eigenvalue $0$ or $1$.
The total fermion number operator is denoted as, $N_f= \sum_i n_i$,
with $N_f=0$ and $N_f=N$ corresponding to the fermionic and the conjugate
vacuum, respectively. The action of $\psi_i, \psi_i^{\dagger}$ on an
arbitrary eigenstate $|n_1, \dots, n_i, \dots, n_N\rangle$ of $N_f$ is the
following:
\bea
\psi_i |n_1, \dots, n_i, \dots, n_N\rangle & = & 0, \ if \ n_i=0\nonumber \\
& = & |n_1, \dots, 0, \dots, n_N\rangle, \ if \ n_i=1\nonumber \\
\psi_i^{\dagger} |n_1, \dots, n_i, \dots, n_N\rangle & = & 0,\ if \
n_i=1\nonumber \\
& = & |n_1, \dots, 1, \dots, n_N\rangle, \ if \ n_i=0.
\eea
\noindent It may be noted that the eigenstate
$|n_1, \dots, n_i, \dots, n_N\rangle$ of $N_f$ can be constructed in the
fermionic Fock space by taking linear superposition of $^NC_{N_f}$ number of
base states:
\be
|i_1, i_2, \dots, i_{N_f} \rangle :=
\psi_{i_1}^{\dagger} \psi_{i_2}^{\dagger} \dots \psi_{i_{N_f}}^{\dagger}
|0\rangle, \ \ i_1 < i_2 < \dots i_{N_f}.
\label{base_fock}
\ee
\noindent The action of the permutation operator on the base states is the
following\cite{rus},
\bea
&& K_{ij} |i_1, \dots, i, \dots, j, \dots, i_{N_f}\rangle =
|i_1, \dots, j, \dots, i, \dots, i_{N_f}\rangle\nonumber \\
&& K_{ij} |i_1, \dots, i, \dots, i_{N_f}\rangle =
|i_1, \dots, j, \dots, i_{N_f}\rangle \ \ j \neq i_1, i_2, \dots
i_{N_f}\nonumber \\
&& K_{ij} |i_1, \dots, i_{N_f}\rangle =
|i_1, \dots, i_{N_f}\rangle \ \ \ i, j \neq i_1, i_2, \dots i_{N_f}.
\eea
\noindent The permutation operator $K_{ij}$ leaves the vacuum of fermionic
Fock space invariant. The operator $K_{ij}$ realizes\cite{rus} a tensor
representation of rank $N_f$ of the symmetric group $S_N$ of permutations
of fermionic operators $\psi_i^{\dagger}$ on the base states
(\ref{base_fock}) with fixed fermion number $N_f$.

An equivalent expression of $\xi_{2N+1}$ in terms of $n_i$ can be written as,
\be
\xi_{2N+1} = (-1)^N \prod_{i=1}^N \left ( 2 n_i - 1 \right ). \ \
\ee
\noindent The action of $\xi_{2N+1}$ on a state $|N_f \rangle$ with fermion
number $N_f$ is,
\be 
\xi_{2N+1} |N_f \rangle = (-1)^{N_f} |N_f \rangle, \ \
0 \leq N_f \leq N.
\ee
\noindent Note that $\xi_{2N+1}$ leaves the fermionic vacuum invariant,
i.e., $\xi_{2N+1} |0 \rangle = |0 \rangle $. On the other hand,
$\xi_{2N+1} |\bar{0} \rangle = (-1)^N |\bar{0} \rangle $, implying that the
conjugate vacuum is invariant for even $N$ and changes sign for odd $N$.

\subsection{Realization of ${\cal{N}}=2$ Super-algebra}

The ${\cal{N}}=2$ superalgebra is realized in terms of the real supercharges
$Q_1$ and $Q_2$ as follows,
\bea
&& Q_1= \frac{1}{\sqrt{2}} \sum_{i=1}^N \left [ \xi_i p_i + \xi_{N+i} W_i
\right ],\nonumber \\
&& Q_2= - \frac{1}{\sqrt{2}} \sum_{i=1}^N \left [ \xi_{N+i} p_i - \xi_{i} W_i
\right ],\ \ \ W_i \equiv \frac{\partial W}{\partial x_i},
\label{realq}
\eea
\noindent where $x_i, p_i=-i \frac{\partial}{\partial x_i}$ are
the position and the momentum operators, respectively. The function
$W(x_1, x_2, \dots, x_N)$ is identified as the superpotential. The
Hamiltonian for the above choices of the supercharges is expressed
as,
\be
H=\frac{1}{2} \sum_{i=1}^N \left (p_i^2 + W_i^2 \right ) - 
\frac{i}{2} \sum_{i,j=1}^N \xi_i \xi_{N+j} W_{ij}, \ \
W_{ij} \equiv \frac{\partial^2 W}{\partial x_i \partial x_j}.
\label{hami}
\ee
\noindent The matrix representation of $\xi_p$'s are taken to be hermitian.
Consequently, the Hamiltonian is hermitian for any real superpotential.
It may be noted that $W_{ij}$ is symmetric with respect to its indexes,
i.e. $W_{ij}=W_{ji}$. Further, the identity $\sum_{i=1}^N W_i=0$ holds
for any transnational invariant superpotential.

It is sometimes convenient to study supersymmetric quantum mechanics
in terms of complex supercharges, 
\bea
Q :=\frac{1}{\sqrt{2}} \left ( Q_1 - i Q_2 \right )
=\sum_{i=1}^N \psi_i^{\dagger} A_i, \ \ A_i:= p_i - i W_i,\nonumber \\
Q^\dagger:=\frac{1}{\sqrt{2}} \left ( Q_1 + i Q_2 \right ) =
\sum_{i=1}^N \psi_i A_i^{\dagger} \ \ A_i^{\dagger} := p_i + i W_i.
\label{compq}
\eea
\noindent The operators $A_i$ and $A_i^{\dagger}$ satisfy the following
relations:
\be
\left [ A_i, A_j \right ]=0=\left [A_i^{\dagger}, A_j^{\dagger} \right ],
\ \ \left [ A_i, A_j^{\dagger} \right ] = \left [ A_j, A_i^{\dagger} \right ]
= 2 W_{ij}.
\ee
\noindent The superalgebra (\ref{salgebra}) is re-written in terms of 
$Q$ and its adjoint $Q^{\dagger}$ as,
\be
H = \frac{1}{2} \left \{ Q, Q^{\dagger} \right \},\ \ \
Q^2=0=(Q^{\dagger})^2, \ \
\left [ H, Q \right ] = 0 = \left [ H, Q^{\dagger} \right ].
\label{c_s_algebra}
\ee
\noindent The supersymmetric Hamiltonian $H$ in Eq. (\ref{hami}) has
the following expression in terms of the fermionic operators:
\bea
H & = & \frac{1}{4} \sum_{i=1}^N \{ A_i, A_i^{\dagger} \} + \frac{1}{4}
\sum_{i,j=1}^N \left [ A_i, A_j^{\dagger} \right ] \left [\psi_i^{\dagger},
\psi_j \right ]\nonumber \\
& = & \frac{1}{2} \sum_{i=1}^N \left ( p_i^2 + W_i^2 - W_{ii} \right )
+ \sum_{i,j=1}^N W_{ij} \psi_i^{\dagger} \psi_j.
\label{eq4}
\eea
\noindent The supersymmetry-preserving phase is characterized by the existence
of state(s) with ground state energy $E_0^s=0$, while that of
supersymmetry-breaking phase as $E_0^s > 0$. The wave-function should be
well-behaved in both the cases.  
The first equation of (\ref{c_s_algebra}) implies that the ground
state of $H$ with $E_0^s=0$ is annihilated by both
$Q$ and $Q^{\dagger}$. The ground states $\Phi_0, \Phi_N$ are determined
from the defining relations of the supercharge in Eq. (\ref{compq}),
$$ \Phi_0 = e^{-W} |0>, \ \ \Phi_N = e^{W} |\bar{0}>.
$$
\noindent Both $\phi_0$ and $\phi_N$ will not be normalizable simultaneously
for the type of superpotential that will be considered in this article.
The supersymmetric phase is characterized by normalizable $\Phi_0$ or/and
$\Phi_N$ for which each $p_i$ is self-adjoint.

The total fermion number operator $N_f$ commutes with the Hamiltonian
and simultaneous eigenstates of $H$ and $N_f$ can be constructed.
The $2^N$ dimensional fermionic Fock space is decomposed into $N+1$
fermionic sectors with $0 \leq N_f \leq N$. The projection of the Hamiltonian
$H$ to a fixed fermionic sector with the fermion number $N_f$ produces
$^N C_{N_f} \times ^N C_{N_f}$ matrix-Hamiltonian $H^{(N_f)}$ and the identity
$\sum_{{N_f}=0}^N {^N C_{N_f}}=2^N$ holds trivially. The projected Hamiltonian
$H^{(N_f)}$ is obtained by evaluating $H$ in the basis given in
Eq. (\ref{base_fock}):
\be
H^{(N_f)} = \langle i_{N_f}, \dots, i_1| H | i_1, \dots, i_{N_f} \rangle.
\ee
In the same basis, the Hamiltonian $H$ has a block-diagonal structure,
$H=diag \{ H^{(N)}, H^{(N-1)}, \dots, H^{(N_f)}, \dots, H^{(1)}, H^{(0)} \}$.
The zero-fermion sector and the $N$-fermion sector of $H$ define
the Hamiltonians,
\bea
H^{(0)} & = & \frac{1}{2} \sum_{i=1}^N \left ( p_i^2 + W_i^2 - W_{ii}
\right ),\nonumber \\
H^{(N)} & = & \frac{1}{2} \sum_{i=1}^N \left ( p_i^2 + W_i^2 + W_{ii}
\right ).
\eea
\noindent If the superpotential depends on an overall multiplicative parameter
$\lambda$, i.e. $ W \sim \lambda W$, then the bosonic potentials of $H^{(0)}$
and $H^{(N)}$ are shape-invariant. In particular, $H^{(N)}(\lambda) =
H^{(0)} (-\lambda)$ for any choice of the superpotential. The bosonic
potentials of these two Hamiltonians may be shape invariant for specific
choices of $W$. However, as Eq. (\ref{intw}) implies, the eigenspectra of
$H^{(0)}$ and $H^{(N)}$ are not in one-to-one correspondence. Thus, the
shape invariance condition can not be implemented as in the case of
system with $N=1$. The $N \times N$ matrix Hamiltonian corresponding to $N_f=1$
has the form,
\be
\left [ H^{(1)} \right ]_{ij} = \delta_{ij} \left ( \frac{1}{2} \sum_{k=1}^N
\left ( p_k^2 + W_k^2 - W_{kk} \right ) \right ) + W_{ij}.
\ee
\noindent The matrix-Hamiltonians $H^{(N_f)}$ for higher $N_f < N$ can be
obtained in a similar way.  However, diagonalizing the Hamiltonian $H^{(N_f)}$
for arbitrary $N_f$ or implementing the shape invariance condition
successfully is a daunting task.

For quantum systems with only one degree of freedom, i.e. $N=1$, the fermionic
Fock space is two dimensional. The Hamiltonian $H$ has a block-diagonal
structure $H=diag \{ H^{(1)}, H^{(0)} \}$, where $H^{(0)}$ and $H^{(1)}$
are identified as partner Hamiltonians\cite{cooper}. In the same fermionic
basis (\ref{base_fock}), the supercharge $Q$ and $Q^{\dagger}$ can be
expressed as $2 \times 2$ matrix operators:
\be
Q = \pmatrix{ {{0}} & {{A_1}}\cr \\
{{0}} & {{0}} }, \ \ \ \ 
Q^{\dagger} = \pmatrix{ {{0}} & {{0}}\cr \\
{{A_1^{\dagger}}} & {{0}} }. \ \ \ \ 
\ee
The superalgebra relates the partner Hamiltonians through the intertwining
relations,
\bea
&& \left [ H, Q \right ] = 0 \  \Rightarrow \ H^{(1)} A_1 = A_1
H^{(0)},\nonumber \\
&& \left [H, Q^{\dagger} \right ] = 0 \ \Rightarrow
H^{(0)} A_1^{\dagger} = A_1^{\dagger} H^{(1)}.
\eea
\noindent These relations allow an algebraic derivation of the complete spectra
and the associated eigen states of the partner Hamiltonians for shape
invariant potentials\cite{cooper}.

The scenario for many-particle systems with $N \geq 2$ is quite different
from the one described above. The supercharge $Q$ ( $Q^{\dagger}$) changes
the fermion number from $N_f$ to $N_f+1(N_f-1)$ and has the following
over-diagonal(under-diagonal) structure\cite{andrianov} in the fermionic
basis (\ref{base_fock}),
\be
\left [ Q \right ]_{pq} = \delta_{p, \ p+1} Q_{p-1,p}, \ \
\left [ Q^{\dagger} \right ]_{pq} = \delta_{p \ p-1} Q_{p,p-1}^{\dagger},
\ \ p, q=1, 2, \dots N_f,
\ee
\noindent where $\delta_{pq}$ is the Kronecker delta,
$Q_{p-1,p}:=\langle i_p,\dots, i_1|Q|i_1, \dots, i_{p-1} \rangle$ is a matrix
of dimension $^NC_{p-1} \times ^NC_p$ depending on the operators $A_i$ and
$Q_{p,p-1}^{\dagger}:=\langle i_{p-1}, \dots, i_1| Q^{\dagger} |
i_1, \dots, i_p\rangle $ is a matrix of dimension $^NC_p \times ^NC_{p-1}$
depending on the operators $A_i^{\dagger}$. The underlying superalgebra leads
to the following intertwining relations\cite{andrianov}:
\bea
&& H^{(N-i)} Q_{i,i+1} = Q_{i,i+1} H^{(N-i-1)},\nonumber \\ 
&& H^{(N-i-1)} Q_{i+1,i}^{\dagger} = Q_{i+1,i}^{\dagger} H^{(N-i)}, \ \
i=0, 1, \dots, N-1.
\label{intw}
\eea
\noindent It follows that a part of the spectrum of $H^{(N_f)}$ coincide with
the spectra of neighbouring block-Hamiltonians $H^{(N_f-1)}$ and $H^{(N_f+1)}$.
However, a successful scheme of implementing the condition of shape invariance
is still beyond the reach for arbitrary $N$. The diagonalization of $H^{(N_f)}$
involves solving a set of $^N C_{N_f}$ coupled second order partial
differential equations which is in general a difficult task.
These technical limitations restrict the number of physically relevant exactly
solved many-particle supersymmetric quantum systems. 

\section{Exactly Solved Systems in One Dimension}

Several supersymmetric many-particle quantum systems may be obtained
by suitably choosing the superpotential $W$. Different choices of the
superpotential lead to different interaction terms in the many-particle
Hamiltonian $H$. The main focus of this review article is on many-body
systems with inverse-square interactions. The bosonic potential
in $H$ scales inverse-squarely if the superpotential is chosen as,
\be
W = W_0 \equiv - ln \ G(x_1, \dots, x_N), \ \ \
\sum_{i=1}^N x_i \frac{\partial G}{\partial x_i} = d,
\label{sp_inv}
\ee
\noindent where $d$ is the degree of the homogeneous function $G$. In general,
Hamiltonians with purely scale-invariant bosonic potentials do not admit
bound states. In the present article, harmonic confining potential will be
added for the description of bound states for which the superpotential is
of the form,
\be
W = W_0 + \frac{\omega}{2} \sum_{i=1}^N x_i^2.
\label{sph}
\ee
\noindent The bosonic potential in $H$ due to the superpotential $W$
in Eq. (\ref{sph}) contains harmonic confining potential and inverse-square
many body interactions. It may be noted that the cross-term arising from
$\sum_{i=1}^N W_i^2$ produces an additive constant in $H$ that is equal to
the degree of the homogeneous function $G$. 

The supersymmetric Hamiltonian $H_0$ corresponding to the superpotential
$W_0$ has $O(2,1)$ symmetry. The Dilatation operator $D$ and the conformal
operator $K$,
\be
D = - \frac{1}{4} \sum_{i=1}^N \left ( x_i p_i + p_i x_i \right ), \ \
K = \frac{r^2}{2}, \ \ r^2 \equiv \sum_{i=1}^N x_i^2,
\label{g021}
\ee
\noindent along with $H_0$ satisfy the $O(2,1)$ algebra (\ref{ap_II.2}).
The Casimir operator $C$ of the $O(2,1)$, as given in
Eq. (\ref{ap_II.3}), has the following expression,
\bea
C & = & \frac{1}{4} \left [ \sum_{i < j=1}^N L_{ij}^2 +
r^2 V + \frac{1}{4} N (N-4) \right ],\nonumber \\
V & \equiv & \sum_{i=1}^N \left [ \left ( \frac{\partial W_0}{\partial x_i}
\right )^2 - \frac{\partial^2 W_0}{\partial x_i^2 } \right ]
+ 2 \sum_{i,j=1}^N \frac{\partial^2 W_0}{\partial x_i \partial x_j} 
\psi_i^{\dagger} \psi_j,
\label{casi}
\eea
\noindent where $L_{ij} \equiv x_i p_j - x_j p_i$ are defined as the angular
momentum operators. The potential $V$ scales inverse-squarely. Consequently,
in the $N$-dimensional hyperspherical coordinate system, $r^2 V$ contains only
angular variables. The Hamiltonian $H=H_0 + \omega^2 K$ thus always can be
separated
into an angular and a radial part in the $N$-dimensional hyperspherical
co-ordinate. The Hamiltonian $H$ contains a harmonic plus an inverse-square
interaction,
\be
H = \frac{p_r^2}{2} + \frac{\omega^2}{2} r^2 + \frac{C^{\prime}}{2 r^2}, \ \
C^{\prime}:= 4 C - \frac{1}{4} N (N-4),
\ee
\noindent with the co-efficient of the inverse-square term determined in
terms of the eigenvalues of the Casimir. The Hamiltonian is integrable for a
fixed eigenvalue of $C$. Infinitely many exact eigenstates and the
corresponding eigenspectra may be obtained analytically for a fixed
eigen value of $C$. However, the eigen value equation for $C$ can be solved
completely for specific superpotentials only and the corresponding potentials
belong to exactly solved systems.

\subsection{Rational $A_{N+1}$ Calogero Model}

There are very few many-particle quantum systems for which the complete
eigen spectra and the associated eigen states can be obtained analytically,
the rational Calogero model being one of them\cite{cs}. The Hamiltonian for
this model describes $N$ particles interacting with each other on a line
through pair-wise inverse-square plus harmonic interactions. The supersymmetric
version of this Hamiltonian was first considered in Ref. \cite{fm} and
was shown to be exactly solvable. The supersymmetric generalizations of
Calogero-Sutherland models based on all the root systems with rational,
trigonometric and hyperbolic potentials have also been
considered\cite{turb,susy,me,sasa,sp0}.
The main emphasis of this topical review being systems with $OSp(2|2)$
supersymmetry, the discussion is restricted to rational models corresponding
to different root systems. The rational $A_{N+1}$ Calogero model is described
below and rational $BC_{N+1}$ model is included in Appendix-IV in section 9.4.

The superpotential for the $A_{N+1}$-type rational Calogero model is given by,
\be
W=- \lambda ln \prod_{i<j} x_{ij} + \frac{\omega}{2} \sum_i x_i^2, \ \
x_{ij}=x_i-x_j.
\label{eq6.1}
\ee
\noindent The Hamiltonian (\ref{eq4}), with the above choice of $W$, has
the following form,
\bea
H & = & - \frac{1}{2} \sum_i \frac{\partial^2}{\partial x_i^2} +
\frac{1}{2} \lambda (\lambda-1) \sum_{i \neq j} x_{ij}^{-2} +
\frac{\omega^2}{2} \sum_i x_i^2 - \frac{\omega}{2}  N \left [  1 +
\lambda \left ( N - 1 \right ) \right ]\nonumber \\
& + & \omega \sum_i \psi_i^{\dagger} \psi_i  +
\lambda \sum_{i \neq j} x_{ij}^{-2} \left ( \psi_i^{\dagger} \psi_i -
\psi_i^{\dagger} {\psi_j} \right ).
\label{eq6.2}
\eea
\noindent  The Hamiltonian is invariant under the permutation symmetry:
\be
x_i \leftrightarrow x_j, \psi_i \leftrightarrow \psi_j, \psi_i^{\dagger}
\leftrightarrow \psi_j^{\dagger}.
\label{permu}
\ee
\noindent The last term in (\ref{eq6.2}) can be expressed in terms of the
fermionic exchange operator $K_{ij}$ defined in Eq. (\ref{fermi_ex}) as,
\be
\sum_{i \neq j} x_{ij}^{-2} K_{ij} .
\ee
\noindent The configuration space of the system is divided into $N!$ different
sectors characterized by a definite ordering of the coordinates of the
particles, $ x_1 < x_2 < \dots < x_N$ and its all possible permutations.
The inverse-square interaction is singular at the coinciding
points $x_i=x_j$. The many-body wave-functions and the associated probability
currents are taken to be vanishing at these points which allows
a smooth continuation of the wave-function from a given sector in the
configuration space to all other sectors.

The ground state wave-function of the super-Hamiltonian with $E_0^s=0$
reads,
\be
\Phi_0 = \phi_0 \ |0\rangle, \ \ \
\phi_0 \equiv \prod_{i < j=1}^N x_{ij}^{\lambda} e^{-\frac{\omega}{2}
\sum_{i=1}^N x_i^2},
\ee
\noindent which is normalizable for $\lambda > - \frac{1}{2}$. However,
a negative value for the parameter in the range $ - \frac{1}{2} < \lambda < 0$
necessarily leads to singularities in $\Phi$ at the coinciding points
$x_i=x_j$. A stronger criteria that each momentum operator $p_i$ is
self-adjoint for the wave-functions of the form $\Phi_0$ requires
$\lambda > 0$. The supersymmetry is preserved for $\lambda > 0$, while
it is broken for $\lambda < 0$. 

\subsection{Shape Invariance and Exact Solvability}

The Hamiltonian in the zero-fermion sector reduces to the rational $A_{N+1}$
Calogero model:
\bea
H^{(0)}(\lambda, \omega) & = & {\cal{H}}^{A_{N+1}} - E_0^{A_{N+1}}\nonumber \\
{\cal{H}}^{A_{N+1}} & := &
- \frac{1}{2} \sum_i \frac{\partial^2}{\partial x_i^2} +
\frac{1}{2} \lambda (\lambda-1) \sum_{i \neq j} x_{ij}^{-2} +
\frac{\omega^2}{2} \sum_i x_i^2,\nonumber \\
E_0^{A_{N+1}} & \equiv & \frac{\omega}{2}  N \left [  1 +
\lambda (N-1) \right ].
\label{H_0_f}
\eea
\noindent The complete eigen values and the eigenstates of $H^{(0)}$ and
${\cal{H}}^{A_{N+1}}$ can be obtained using the ideas of supersymmetry and
shape invariance\cite{me}. It may be noted that many-body potentials of
the Hamiltonians $H^{(0)}$ and $H^{(N)}$ are shape-invariant,
\bea
H^{(N)}(\lambda, \omega) & = & H^{(0)}(\lambda+1, \omega)
+ \frac{\omega}{2} N (N+1)\nonumber \\
& = & - \frac{1}{2} \sum_i \frac{\partial^2}{\partial x_i^2} +
\frac{1}{2} \lambda (\lambda + 1) \sum_{i \neq j} x_{ij}^{-2} +
\frac{\omega^2}{2} \sum_i x_i^2\nonumber \\
& + & \frac{\omega N}{2} \left [  1 - \lambda (N-1) \right ].
\label{H_N_f}
\eea
\noindent However, no direct relation between the eigen-spectra of
$H^{(0)}$ and $H^{(N)}$ can be shown, as in the case of supersymmetric
quantum mechanics with one bosonic and one fermionic degrees of freedom.
Nevertheless, the shape invariance can be used to obtain the spectrum of
$H^{(0)}$ or $H^{(N)}$ with the introduction of permutation and Dunkl
operators\cite{me}.

The Dunkl operator for the rational $A_{N+1}$ Calogero model is defined as,
\be
\pi_i = p_i + i \lambda \sum_{j(\neq i)} x_{ij}^{-1} M_{ij},
\ee
\noindent where the exchange operator $M_{ij}$ satisfies the following
properties\cite{poly,revpoly},
\bea
&& M_{ij} = M_{ij}^{-1} = M_{ij}^{\dagger} = M_{ji}, \ \
M_{ij} \phi^{\pm} = \pm \phi^{\pm},\nonumber  \\
&& M_{ij} O_i = O_j M_{ij}, \ \ M_{ij} O_k = O_k M_{ij} \ \ \ if \ \ i, j, k \
\ distinct,
\nonumber\\
&& M_{ijk}:= M_{ij} M_{jk}, \ \ M_{ijk}=M_{jki}=M_{kij}.
\eea
\noindent The function $\phi^{+}(\phi^{-})$ appearing above is a(an)
symmetric(anti-symmetric) function of the $N$ bosonic co-ordinates and $O_i$
is a single-particle bosonic operator in the phase space. The exchange operator
$M_{ij}$ acting on any fermionic operator leaves it unchanged. The Dunkl
operators $\pi_i$ commute with each other, i.e. $\left [ \pi_i, \pi_j \right ]
=0$. A set of operators $a_i, a_i^{\dagger}$ are introduced,
\be
a_i:= \pi_i - i \omega x_i, \ \ a_i^{\dagger} := \pi_i + i \omega x_i,
\ee
\noindent which satisfy an extended version of the Heisenberg algebra involving
the permutation operators $M_{ij}$:
\be
\left [ a_i, a_j^{\dagger} \right ] = 2 \omega \delta_{ij} \left ( 1 +
\lambda \sum_{k(\neq i)} M_{ik} \right ) - 2 \left ( 1 - \delta_{ij} \right )
\lambda \omega M_{ij}.
\ee
\noindent All other commutators involving $a_i$ and their adjoints vanish
identically. The Hamiltonian ${\cal{H}}$ and its partner Hamiltonian
$\tilde{{\cal{H}}}$ are introduced as follows:
\be
{\cal{H}} = \frac{1}{2} \sum_{i=1}^N a_i^{\dagger} a_i, \ \
\tilde{{\cal{H}}} = \frac{1}{2} \sum_{i=1}^N a_i a_i^{\dagger}.
\ee
\noindent The Hamiltonian ${\cal{H}}$ reduces to $H^{(0)}$ if $M_{ij}$ acts on
symmetric functions only, whereas $\tilde{{\cal{H}}}$ becomes $H^{(N)}$
if $M_{ij}$ acts on antisymmetric functions only. No choice between the
symmetric and the antisymmetric functions will be made in the discussions
below and the results obtained are valid for either cases. The convention
is that an(a) upper(lower) sign in the expressions below corresponds to
the case that $M_{ij}$ only acts on symmetric(antisymmetric) functions.

The extended Heisenberg algebra can be used to show one to one correspondence
between the non-zero energy eigen values of ${\cal{H}}$ and $\tilde {\cal{H}}$.
In particular, if $\phi$ is the eigenstate of ${\cal{H}}$ with eigenvalue
$E (>0)$, then $A_1 \phi$ is the eigenstate of $\tilde{\cal{H}}$ with
eigenvalue $E+ \delta_1$ i.e.
\be\label{16}
\tilde{\cal{H}} (A_1 \phi)=[E+ \delta_1](A_1 \phi),
\ee
where the operator $A_1$ is defined as,
\be\label{ani1}
A_1 :={\sum_{i}} a_{i}, \ \ \
\delta_1 = \left [ \left ( N-1 \right ) \pm \lambda  N(N-1) \right ] \omega.
\ee
\noindent Similarly, if $\tilde{\phi}$ is the eigen state of $\tilde{\cal{H}}$
with eigenvalue $\tilde E$, then $A_1^{\dagger} \phi$ is the eigenfunction of
${\cal{H}}$ with eigenvalue $\tilde{E} - \delta_1$ i.e.
\be\label{19}
{\cal{H}} ({{A}_1^\dagger}\tilde {\phi}) = \left [ \tilde {E}-\delta_1 \right ]
\left ( {{A}_1^\dagger} \tilde{\psi} \right ).
\ee
\noindent  The energy eigenvalues and eigenfunctions of the two partner
Hamiltonians ${\cal{H}}$ and $\tilde{\cal{H}}$ are thus related through
the following relations:
\bea\label{20}
&& \tilde{E}_n = E_{n+1} +\delta_1, \ \ E_0 = 0, \ \ n = 0,1,2,\dots\nonumber \\
&& \tilde{\phi}_n = {A_1 \phi_{n+1} \over \sqrt {E_{n+1} +\delta_1}}~~,
~\phi_{n+1} = {A_1^{\dagger} \tilde{\phi}_n \over \sqrt{E_{n+1}}}~~.
\eea
\noindent The standard results of supersymmetric quantum systems with
one bosonic and one fermionic degrees of freedom are reproduced for $N=1$.
The energy levels of ${\cal{H}}$ and $\tilde{\cal{H}}$ are non-degenerate
for $\delta \neq 0$.

The shape invariance condition for the partner Hamiltonians ${\cal{H}}$ and
$\tilde{\cal{H}}$ reads,
\bea\label{23}
&& \tilde{\cal{H}} (\lambda)  = {\cal{H}} (\lambda) + R(\lambda),\nonumber \\
&& R(\lambda) = [ N \pm \lambda N(N-1)] \omega = \omega + \delta_1~~.
\eea
\noindent Following the standard formalism of supersymmetric quantum
mechanics\cite{cooper} and using the first equation of (\ref{20}),
the spectrum of ${\cal{H}}$ is determined\cite{me}:
\bea \label{25}
{E_{n}} & = & n \left ( R(\lambda) - \delta_1 \right )\nonumber \\
&  = & n \omega ~.
\eea
\noindent The eigen values of ${\cal{H}}^{A_{N+1}}$ are thus equal to the
eigen values of $N$ free harmonic oscillators shifted by a constant,
\be
E_n^{A_{N+1}} = E_n + E_0^{A_{N+1}} = \omega n +
\frac{\omega}{2}  N \left [  1 + \lambda (N-1) \right ].
\ee
\noindent The $n^{th}$ eigenstate is obtained as,
$\phi_n =  \left ( A_1^{\dagger} \right )^n \phi_0$, since $A_1$ and
$A_1^{\dagger}$ can be identified as the annihilation and creation operators,
respectively. In general, the following identities hold true,
\bea
&& A_n:= \sum_{i=1}^N a_i^n, \ \
A_n^{\dagger} := \sum_{i=1}^N \left ( a_i^{\dagger}\right )^n, \ n \leq N,
\nonumber \\ 
&& \left [ {\cal{H}}, A_n \right] = - n A_n, \ \
\left [ {\cal{H}}, A_n^{\dagger} \right ] = n A_n^{\dagger}.
\eea
\noindent Further, the operators $A_n, A_n^{\dagger}, {\cal{H}},
{\tilde{\cal{H}}}$ satisfy relations which are analogous to those given by
Eqs. (\ref{16}) and (\ref{19}). This facilitates a construction of all the
degenerate states $\phi_{\{n_i\}}$, 
\be\label{28}
\phi_{\{n_i\}} = \prod_{i=1}^N \left ( {A}^\dagger_i \right )^{n_i} \phi_0,
\ \ \ \ k=\sum_{i=1}^N n_i ~,
\ee
\noindent corresponding to a particular value of $k$. All the corresponding
states of $\tilde{\cal{H}}$ can be obtained by applying the same $A_1$ on
$\phi_{\{n_i\}}$.

\section{$OSp(2|2)$ Supersymmetric Systems}

The supersymmetric rational $A_{N+1}$ Calogero model has a dynamical
$OSp(2|2)$ symmetry\cite{fm}. The celebrated Calogero model is obtained by
projecting
the supersymmetric Hamiltonian $H$ in the zero fermion sector, i.e. $N_f=0$.
There exists a possibility of constructing a new supersymmetric many-particle
quantum system with dynamical $OSp(2|2)$ supersymmetry that is different
from $H$, yet it reduces to the celebrated Calogero model in appropriate limit.
In particular, the Hamiltonian $H$ constructed by Freedman and Mende\cite{fm}
corresponds to `typical' representation of $OSp(2|2)$. The `atypical'
representation of $OSp(2|2)$ with the same superpotential $W$ in (\ref{eq6.1})
produces a many-particle supersymmetric quantum system that is different from
$H$\cite{nonu}. Thus, the construction of $OSp(2|2)$ supersymmetric
Calogero model is not unique.

The structure equations of $OSp(2|2)$ symmetry are given in Appendix-II.
In this section, the superpotential $W$ is replaced by $W_0$ as defined in Eq.
(\ref{sp_inv}) and the corresponding Hamiltonian $H$ in Eq. (\ref{hami})
or in Eq. (\ref{eq4}) is denoted as $H_0$. A coordinate realization of
the dilatation operator $D$ and the conformal generator $K$ is given in
Eq. (\ref{g021}). The real supercharges $Q_1, Q_2$ are described in Eq.
(\ref{realq}). The remaining three generators $S_1, S_2$ and $Y$ corresponding
to the `typical' representation of $OSp(2|2)$ are realized in the following way:
\bea
&& S_1 = \frac{1}{\sqrt{2}} \sum_{i=1}^N \xi_i x_i, \ \
S_2 = -\frac{1}{\sqrt{2}} \sum_{i=1}^N \xi_{N+i} x_i,\nonumber \\
&& Y = - \frac{1}{4} \left ( i \sum_{i=1}^N \xi_{N+i} \xi_i + 2 d \right ).
\label{noo}
\eea
\noindent These operators can be expressed in terms of fermionic operators,
\bea
S & := & \frac{1}{\sqrt{2}} \left ( S_1 - i S_2 \right ) =
\sum_{i=1}^N \psi_i^{\dagger} x_i, \ \
S^{\dagger} := \frac{1}{\sqrt{2}} \left ( S_1 + i S_2 \right ) =
\sum_{i=1}^N \psi_i x_i, \nonumber \\
Y & = & -\frac{N_f}{2} + \frac{N}{4} - \frac{d}{2}.
\eea
\noindent The hypercharge $Y$ is related to the total fermion number
operator $N_f$ and $S$ factorizes the conformal generator $K$.

The Hamiltonian $H_0$ is scale invariant and its ground state
wave-function,
\be
\phi_0 = G(x_1, \dots, x_N) \ |0\rangle,
\ee
\noindent is not even plane-wave normalizable. The time-evolution of 
such scale invariant supersymmetric systems are generally studied
in terms of the operators\cite{dff,fr,nonu},
\be
{\cal{H}}_{\pm} = R \pm \omega Y,
\ \ R:=\frac{1}{2} \left ( H_0 + \omega^2 K \right ).
\label{hami_atypical}
\ee
\noindent The generator of the compact rotation $R$ of $O(2,1)$ and the
hypercharge $Y$ are simultaneously diagonal in the Cartan basis. The operator
${\cal{H}}_+$ is related to ${\cal{H}}_-$ and the vice verse through an
automorphism of the $OSp(2|2)$ algebra. Thus, either ${\cal{H}}_+$ or
${\cal{H}}_-$ may be identified as the supersymmetric Hamiltonian and the
choice in this article is the Hamiltonian ${\cal{H}}_{-}$. The Hamiltonian
${\cal{H}}_-$ can be expressed in terms of supercharges $F$ and $F^{\dagger}$
as follows,
\be
{\cal{F}} := \frac{1}{\sqrt{2}} \left ( Q - i \omega S \right ), \ \
{\cal{F}}^{\dagger} :=\frac{1}{\sqrt{2}} \left ( Q^{\dagger} +
i \omega S^{\dagger} \right ), \ \
{\cal{H}}_- = \frac{1}{2} \left \{ {\cal{F}}, {\cal{F}}^{\dagger} \right \}.
\ee
\noindent The Hamiltonian $H$ in Eqs. (\ref{hami}, \ref{eq4}) with the
superpotential as given in Eq. (\ref{sph}) is identical with $2 {\cal{H}}_-$.

The bosonic generators $H_0, D, K, Y$ and the fermionic generators
$Q_1, Q_2, S_1, S_2$ with the co-ordinate realization described
above satisfy the structure equations of $OSp(2|2)$. The quadratic and
the cubic Casimir operators of $OSp(2|2)$ are non-vanishing
and hence, this particular co-ordinate realization of generators corresponds to
the `typical' representation of $OSp(2|2)$\cite{nonu}. The arguments in
favour of non-vanishing $C_2$ and $C_3$ are as follows. The terms appearing
in the expressions of $C_2$ and $C_3$ in Eq. (\ref{casi1}) have the following
co-ordinate realization:
\bea
i [Q_1, S_1] & = & \frac{N}{2} - \frac{i}{2} \sum_{i,j=1}^N \xi_i \
\xi_j \ L_{ij} + i \sum_{i,j=1}^N \xi_{N+i} \ \xi_j \ x_j \ W_i,\nonumber \\
i [Q_2, S_2] &  =  & \frac{N}{2} - \frac{i}{2} \sum_{i,j=1}^N \xi_{N+i} \
\xi_{N+j} \ L_{ij} - i \sum_{i,j=1}^N \xi_{N+i} \ \xi_j \ x_i \ W_j,\nonumber \\
Y^2 & = & \frac{1}{16} \left ( 4 d^2 + N + i 4 d \sum_i \xi_{N+i} \xi_i
+ \frac{1}{2} \sum_{i \neq j} \left [ \xi_{N+i},
\xi_{N+j} \right ] \xi_i \xi_j \right ).
\label{wonda}
\eea
\noindent The operator $i [Q_1,S_1] + i [Q_2,S_2] - Y^2$ appearing in $C_2$
does not contain any term proportional to $L_{ij}^2$. However, the Casimir
operator $C$ in Eq. (\ref{casi}) contains a term proportional to $ L_{ij}^2$.
This implies that $C_2$ in (\ref{casi1}) can not vanish identically.
The cubic Casimir operator $C_3$ is also non-vanishing, since it
contains a term of the form $ L_{ij}^2 Y$ which can not be canceled
from rest of the terms appearing in its definition in Eq. (\ref{casi1}).

\subsection{Non-uniqueness of the Construction}

The operators $H_0, D, K, Q_1$ and $S_1$ have identical coordinate realizations
in `typical' as well as in `atypical' representations of $OSp(2|2)$. The
operators $Q_2, S_2$ and $Y$ have different coordinate realizations
corresponding to two different representations of the group. The operators
$Q_2, S_2$ and $Y$ in the `atypical' representation of $OSp(2|2)$ are realized
in the following way,
\bea
&& \hat{Q}_2 = - i \xi_{2N+1} Q_1, \ \
\hat{S}_2 = - i \xi_{2N+1} S_1\nonumber \\
&& \hat{Y} = \frac{\xi_{2N+1}}{2}
\left [ - \frac{i}{2} \sum_{i,j=1}^N \xi_i \xi_j
L_{ij} + i \sum_{i,j=1}^N \xi_{N+i} \xi_j W_i x_j +
\frac{N}{2} \right ].
\eea
\noindent The structure equations of $OSp(2|2)$ are now satisfied by
the bosonic generators $H_0, D, K, \hat{Y}$ and the fermionic generators
$Q_1, \hat{Q}_2, S_1, \hat{S}_2$. The `atypical' realization of the $OSp(2|2)$
superalgebra is possible only for $ N \geq 2$. No independent coordinate
realization of $Q_2, S_2$ and $Y$, other than the one given in Eqs.
(\ref{realq},\ref{noo}), is admissible for systems with one bosonic and
fermionic degrees of freedom. 

The cubic Casimir operator can be expressed in terms of the quadratic Casimir
operator in the `atypical' representation\cite{dhowker}:
\be
C_3 = \left ( \hat{Y} - \frac{\xi_{2N+1}}{4} \right ) C_2.
\label{atypi_c3}
\ee
\noindent Further, the Scasimir operators $C_s, \bar{C}_s$ defined in
Eqs. (\ref{s-casi}) and (\ref{bar_s-casi}) are identical in this
representation. Using the second equation of (\ref{ccseq}), definition
of $C_s$ and the identity,
\be
\hat{Y}^2 = \frac{1}{4} \left ( C_s^2 + C_s + \frac{1}{4} \right ),
\ee
\noindent it follows that the quadratic Casimir operator $C_2$
vanishes identically. Consequently,  the cubic Casimir $C_3$ is identically
equal to zero for this particular representation. The spectrum is not
completely specified by the eigenvalues of Casimir operators and hence, the
representation is `atypical'. The Scasimir $C_s$ can be used to determine
the spectrum, since it commutes with the bosonic generators and anti-commutes
with the fermionic generators.

A new supersymmetric Hamiltonian preserving the $OSp(2|2)$ symmetry
may be introduced in the `atypical' representation:
\bea
\hat{\cal{H}}_{\pm} & = & \frac{1}{2} \left ( H_0 + \omega^2 K \right ) \pm 
\omega \hat{Y}\nonumber \\
& = & \frac{1}{4} \sum_i \left ( p_i^2 + W_i^2 - W_{ii} +
\omega^2 x_i^2 \right ) + \frac{1}{4} \sum_{i,j} \psi_i^{\dagger} \psi_j W_{ij}
\pm \frac{\omega \xi_{2N+1}}{4} {\Big [}  N + 2 d \nonumber \\
& - & i \sum_{i,j} \left ( \psi_i^{\dagger} \psi_j^{\dagger} +
\psi_i^{\dagger} \psi_j \right ) e^W L_{ij} e^{-W} - i
\sum_{i,j} \left ( \psi_i \psi_j -
\psi_j^{\dagger} \psi_i  \right)  e^{-W} L_{ij} e^{W}\nonumber \\
& - & \sum_{i,j} \left ( \psi_i^{\dagger} \psi_j + \psi_j^{\dagger}
\psi_i \right ) \left ( x_i W_j + x_j W_i \right ) {\Big ]}.
\label{h2}
\eea
\noindent where $W$ is identified with $W_0$ defined in Eq. (\ref{sp_inv}).
The Hamiltonian ${\cal{H}}_{\pm}$ can also be casted in a manifestly
supersymmetric form,
\bea
\hat{\cal{H}}_{\pm}  & = & \frac{1}{2} \{ \hat{\cal{F}},
\hat{\cal{F}}^{\dagger} \}\nonumber \\
{\hat{\cal{F}}} & := & \xi_{2N+1}^- \left ( Q_1 - i \omega S_1 \right ), \ \
{\hat{\cal{F}}}^{\dagger} := \xi_{2N+1}^+ \left ( Q_1 +
i \omega S_1 \right ).
\eea
\noindent The difference between $\cal{H}_{\pm}$ and $\hat{\cal{H}}_{\pm}$
lies in the expressions of the hyper-charges $Y$ and $\hat{Y}$, respectively.
Unlike in the case of $\cal{H}_{\pm}$, the total fermion number operator
$N_f$ does not commute with $\hat{\cal{H}}_{\pm}$. Consequently, the
eigenstates of $\hat{\cal{H}}_{\pm}$ can not be constructed as simultaneous
eigenstates of $N_f$. However, $\hat{\cal{H}}_{\pm}$ commutes with $\hat{Y}$
and simultaneous eigenstates of these two operators can be constructed.
If the eigenvalue of $\hat{Y}$ is chosen as ${\cal{Y}}^+= \frac{\omega}{4}
\left ( N + 2 d \right )$ and $R$ is projected in the fermionic vacuum
$|0\rangle$, then $ 2 {\hat{\cal{H}}}_-$ reduces to the purely bosonic
Hamiltonian $H^{(0)}$. Similarly, $ 2 {\hat{\cal{H}}}_-$ reduces to the purely
bosonic Hamiltonian $H^{(N)}$ for the choice of the eigenvalue $\hat{Y}$
as ${\cal{Y}}^-= \frac{\omega}{4} \left ( N - 2 d \right )$ and projection
of $R$ in the conjugate fermionic vacuum $|\bar{0}\rangle$. Thus, the same
Hamiltonians $H^{(0)} ( or H^{(N)})$ may be obtained from two different
supersymmetric Hamiltonians with $OSp(2|2)$ symmetry. 

The Hamiltonian for the rational $A_{N+1}$ Calogero model is given by,
\bea
\hat{\cal{H}}_{\pm} & = & \frac{1}{4} \sum_i \left ( p_i^2 + \omega^2 x_i^2 +
\sum_{j \neq i} \frac{\lambda (\lambda-1)}{(x_i - x_j )^2} \right )
+ \frac{\lambda}{2} \sum_{i\neq j} \frac{\psi_i^{\dagger} \psi_i -
\psi_j^{\dagger} \psi_i}{( x_i - x_j )^2}\nonumber \\
& \pm & \frac{\omega \xi_{2N+1}}{4} {\Big [}  N + \lambda N (N-1)
 - i \sum_{i,j} \left ( \psi_i^{\dagger} \psi_j^{\dagger}
 + \psi_i^{\dagger} \psi_j \right ) \left ( G L_{ij} \
G^{-1} \right )\nonumber \\
& - & i \sum_{i,j} \left ( \psi_i \psi_j - \psi_j^{\dagger} \psi_i  \right)
\left ( G^{-1} L_{ij} \ G\right )\nonumber \\
& - & \lambda \sum_{i,j} \left ( \psi_i^{\dagger} \psi_j + \psi_j^{\dagger}
\psi_i \right ) \left ( \sum_{k\neq j} \frac{x_i}{ x_j - x_k}
+ \sum_{k \neq i} \frac{x_j}{x_i - x_k } \right )  {\Big ]},
\eea
\noindent where $G=\prod_{i <j} x_{ij}^{\lambda}$.
It is not known whether $\hat{\cal{H}}_{\pm}$ is integrable or not.
Many exact eigenstates
of $\hat{\cal{H}}_{\pm}$ can be constructed explicitly using the underlying
$OSp(2|2)$ supersymmetry. The quadratic and the cubic Casimir operators
being zero in the `atypical' representation, these operators can not be
used to characterize the spectra. Further, as in the case of `typical'
Calogero model, $OSp(2|2)$ is not the full spectrum generating algebra
of $\hat{\cal{H}}_{\pm}$. Alternative methods are required to establish
(non-)integrability of $\hat{\cal{H}}_{\pm}$.

\subsection{Mapping to Free Super-oscillators}

The Hamiltonian $H$ in Eq. (\ref{eq6.2}) is exactly solvable in both
supersymmetry-preserving and supersymmetry-breaking phases\cite{fm,bh,turb}.
The spectrum in the supersymmetry-preserving phase is identical to that
of the $N$ free super-oscillators, while in the supersymmetry-breaking phase,
it has no counter-part in the super-oscillator model\cite{fm}. This is
primarily because of the fact that the supersymmetry is always preserved
in the super-oscillator model, once the convention for choosing the ground
state in either zero or $N$ fermion sector has been made.
This is a good indication that the rational $A_{N+1}$ Calogero model in the
supersymmetry-preserving phase may be mapped to a set of free superoscillators
through a similarity transformation, much akin to its non-supersymmetric
version\cite{pani}. The superpotential of the rational $A_{N+1}$ Calogero
model is given in Eq. (\ref{eq6.1}) and without any loss of generality
$\omega$ is taken to be unity in this section.

\subsubsection{Supersymmetry-preserving Phase}

The mapping of $H$ to a system of free super-oscillators is achieved as follows:
\bea
H_1 & = & e^{W} H e^{-W} \\
& = & \sum_i \left ( x_i \frac{\partial}{\partial x_i} +
\psi_i^{\dagger} \psi_i \right ) - S ,\nonumber \\
S & := & \frac{1}{2} \sum_i \frac{\partial^2}{\partial x_i^2} +
\lambda \sum_{i \neq j} x_{ij}^{-1} \frac{\partial}{\partial x_i}
 - \lambda \sum_{i \neq j} x_{ij}^{-2} \left ( \psi_i^{\dagger} \psi_i -
\psi_i^{\dagger} \psi_j \right ).
\label{lasalle}
\eea
\noindent It may be noted that $N_f$ commutes with both $H_1$ and $S$.
Making use of the following identity,
\be
\left [ \sum_i \left ( x_i \frac{\partial}{\partial x_i} +
\psi_i^{\dagger} \psi_i \right ) , S \right ] = - 2 S,
\ee
\noindent the operator $H_1$ is mapped to the operator $H_2$,
\bea
&& H_2 = e^{\frac{S}{2}} H_1 e^{-\frac{S}{2}}\nonumber \\
&& \ \ \ \ = T H T^{-1}\nonumber \\
&& \ \ \ \ = \sum_i \left ( x_i \frac{\partial}{\partial x_i} +
\psi_i^{\dagger} \psi_i \right ), \ \
T:=e^\frac{S}{2} e^W,
\label{map_osc}
\eea
\noindent which may be considered as a supersymmetric generalization of
the Euler operator. The familiar form of the super-oscillator Hamiltonian
may be obtained in the following way,
\bea
H_{sho} & = & e^{-\frac{1}{2} \sum_i x_i^2 } e^{-
\frac{1}{4} \sum_i \frac{\partial^2}{\partial x_i^2}} H_2
e^{\frac{1}{4} \sum_i \frac{\partial^2}{\partial x_i^2}}
e^{\frac{1}{2} \sum_i x_i^2 }\nonumber \\
& = &  \frac{1}{2} \sum_i \left ( - \frac{\partial^2}{\partial x_i^2}
+ x_i^2 \right ) + \sum_i \psi_i^{\dagger} \psi_i -
\frac{N}{2}.
\label{eq6.7}
\eea
\noindent The mapping of $H$ to $H_{sho}$ acts as a necessary condition
for establishing the equivalence between the two. The sufficient condition
is satisfied provided the complete spectrum of $H$ is obtained from that
of $H_2$ or $H_{sho}$ by using the similarity operator $T$. Thus, the
similarity operator $T$ should not take the original Hamiltonian out of its
Hilbert space and the domain of $T$, $H$ and $H_2(H_{sho})$ should be identical.

If $P_{n,k}$ is an eigen-function of $H_2$ with the eigen-value
$E_{n,k}$, then, $H$ has the same eigen-value $E_{n,k}$ with the
eigen-function given by,
\be
\chi= T^{-1} P_{n,k} \ |0> .
\ee
The eigenfunctions $P_{n,k}$, which are invariant under the combined
exchange of the bosonic and the fermionic coordinates, i.e.
$(x_i, \psi_i) \leftrightarrow (x_j, \psi_j)$, produce physically
acceptable $\chi$. Any departure from this prescription for choosing
$P_{n,k}$ produces essential singularity in $\chi$ and is not physically
acceptable. The complete eigenstates and eigen spectrum of $H$ can be
reproduced using the similarity transformation and a complete set of
$P_{n,k}$ that is constructed using the prescription described above.
This shows the equivalence between $H$ and $H_{sho}$. An example of
$P_{n,k}$ that corresponds to $N_f=1$ solution of $H_2$ with energy
eigenvalue $E_{n,k}= 2n+k$ is given as,
\be
P_{n,k}= r^{2 n} \sum_{i=1}^N x_i^{k-1} \psi_i^{\dagger}, \ \
n=0, 1, \dots, k=1, 2, \dots.
\ee
\noindent The action of $S^m, m \geq 1$ on $P_{n,k}$ does not produce
any singularity\cite{susy}. In general, $P_{n,k}$ for arbitrary $N_f$
may be chosen as,
\be
P_{n,k} = \frac{1}{N_f!} r^{2 n}
\sum_{i_1, i_2, \dots, i_{N_f}} f_{i_{1} i_{2} \dots i_{N_f}}
(x_1, x_2, \dots, x_N) \  \psi_{i_1}^{\dagger}
\psi_{i_2}^{\dagger} \dots \psi_{i_{N_f}}^{\dagger},
\label{eq6.10.1}
\ee
\noindent where $ f_{i_{1} i_{2} \dots i_{N_f}}$ is anti-symmetric under
the exchange of any two indices and is a homogeneous function of degree
$k-N_f$. The anti-symmetric nature of $f$ ensures that $P_{n,k}$ is
permutation invariant under the combined exchange of bosonic and
fermionic coordinates. No closed form expression of 
$f_{i_{1} i_{2} \dots i_{N_f}}$ for arbitrary $N_f$ is known.

An algebraic construction of the eigen spectrum of $H$ is allowed through
the introduction of the operators\cite{susy},
\bea
&& b_i^{-}= i p_i, \ \ b_i^{+}= 2 x_i, \ \ \psi_i^+:=\psi_i^{\dagger}, \ \
\psi_i^-:= \psi_i\nonumber \\
&& B_n^{\pm}=  T^{-1} \left ( \sum_{i=1}^N b_i^{{\pm}^n} \right ) T,\ \
F_n^{\pm}=T^{-1} \left ( \sum_i \psi_i^{\pm} b_i^{{\pm}^{n-1}}
\right ) T,\nonumber \\
&& q_n^{\pm} = T^{-1} \left ( \sum_{i} \psi_i^{\mp} \left
( b_i^{\pm} \right )^n \right ) T. 
\eea
\noindent The operators 
$B_n^{\pm}$ and $F_n^{\pm}$ satisfy the algebra of $N$ independent
superoscillators with frequencies $1,2, \dots, N$, namely,
\bea
&& [H,B_n^{\pm}]=\pm n B_n^{\pm}, \ \ [H, F_n^{\pm}]=\pm n
F_n^{\pm},\nonumber \\
&& \left \{ q_1^+, F_n^+ \right \} = B_n^+, \ \
\left [ q_1^-, B_n^+ \right ] = 2 n F_n^+,
\label{alg_osc}
\eea
\noindent and so on. Thus,
\be
\chi_{n_1 \dots n_{N} \nu_1 \dots \nu_N}= \prod_{k=1}^N B_k^{+^{n_k}}
F_k^{+^{\nu_k}} \ \Phi_0,
\ee
\noindent is the eigenfunction of $H$ with the eigen-value,
\be
E= \sum_{k=1}^{N} k (n_k + \nu_k), \ \ n_k=0, 1, \dots; \nu_k=0,1.
\ee
\noindent The spectrum of $H$ is identical, including degeneracies at each
level, to that of $N$ superoscillators with frequencies $1, 2, \dots,N$.

\subsubsection{Supersymmetry-breaking Phase}

The eigen-spectrum of the rational $A_{N+1}$ Calogero model in the
supersymmetry-breaking phase can also be constructed from the known
super-oscillator basis by making use of a duality property of the
model\cite{susy}. In particular, a new super-Hamiltonian $H_d$ may be
constructed\cite{fm} from $H$ by using the transformation:
$\lambda \rightarrow - \lambda$ and $\psi_i \leftrightarrow \psi_i^{\dagger}$,
\bea
H_d & = &  - \frac{1}{2} \sum_i \frac{\partial^2}{\partial x_i^2} +
\frac{1}{2} \lambda (\lambda-1) \sum_{i \neq j} x_{ij}^{-2} +
\frac{1}{2} \sum_i x_i^2 + \frac{1}{2}  N \left [  1 +
\lambda \left ( N - 1 \right ) \right ]\nonumber \\
& - &\sum_i \psi_i^{\dagger} \psi_i  +
\lambda \sum_{i \neq j} x_{ij}^{-2} \left ( \psi_i^{\dagger} \psi_i -
\psi_i^{\dagger} {\psi_j} \right ),\nonumber \\
H & = & H_d + 2 N_f - N \left [ 1 + \lambda \left ( N - 1 \right ) \right ].
\label{eq6.14}
\eea
\noindent The ground state of $H_d$ is in the $N$ fermion sector,
\be
{\tilde{\Phi}}= e^{-\tilde{W}} \ |\bar{0}>=
\prod_{i <j} x_{ij}^{-\lambda} e^{-\frac{1}{2} \sum_i x_i^2} \ |\bar{0}>,
\ \ \lambda < 0.
\label{eq6.15}
\ee
\noindent The supersymmetric phase of $H_d$ is described by $\lambda < 0$.
This condition on $\lambda$ ensures that each momentum operator $p_i$ is
self-adjoint for wave-functions of the form $\tilde{\Phi}$.
The Hamiltonian $H_d$ differs from the original Hamiltonian $H$ by
the fermionic number operator and a constant. This implies
that any eigen-function of this dual model is also a valid eigen-function of
the rational $A_{N+1}$ Calogero model. Of course, the corresponding energy
eigen-values are different from each other. For example, the wave-function
$\tilde{\Phi}$ is also an eigen-state of $H$ with positive energy.
This is, in fact, the ground state of $H$ in the supersymmetry-breaking
phase\cite{fm}. The complete spectrum of $H$ in this phase can be obtained
from $H_d$ by making use of the second equation of (\ref{eq6.14}).

The dual Hamiltonian can be shown to be equivalent to a free super-oscillator
Hamiltonian through a similarity transformation\cite{susy}. The eigen-spectrum
thus obtained from the super-oscillator model via the dual Hamiltonian indeed
correctly describes the supersymmetry-breaking phase of the rational $A_{N+1}$
Calogero model. An algebraic construction of the complete set of eigenstates
is admissible with the introduction of the bosonic creation operator
$\hat{B}_n^+$ and the fermionic creation operator $\hat{F}_n^+$,
\bea
&& \hat{B}_n^+ = \sum_i \hat{T}^{-1} b_i^{+^n} \hat{T}, \ \
\hat{F}_n^{+} = \hat{T}^{-1} \left (\sum_i
\psi_i b_i^{+^{n-1}} \right ) \hat{T},\nonumber \\
&& \hat{T} := e^{\frac{S}{2}} e^{\hat{W}}, \ \
\hat{W} := \lambda \ ln \prod_{i <j} x_{ij} + \frac{1}{2} \sum_{i} x_i^2.
\label{eq6.17.1}
\eea
\noindent The eigenstates and the associated eigen values are,
\bea
&& \hat{\Phi}_{n_1, \dots, n_N, \nu_1, \dots, n_N} =
\prod_{k=1}^N \hat{B}_k^{+^{n_k}} \hat{F}_k^{+^{\nu_k}}
\tilde{\Phi},\nonumber \\
&& E=N \left [  1 - \lambda \left ( N - 1 \right ) \right ] +
\sum_{k=1}^N \left [ k n_k + \left ( k - 2 \right ) n_k \right ].
\eea
\noindent The bosonic quantum numbers $n_k$'s
are non-negative integers, while the fermionic quantum numbers $\nu_k$'s
are either $0$ or $1$.

\subsubsection{Generalization}

The mapping of $H$ in Eq. (\ref{eq4}) to a set of free superoscillators
through a similarity transformation is valid for a class of superpotential of
the form given in Eq. (\ref{sph})\cite{susy}:
\bea
&& H_2 = {\tilde{T}} H {\tilde{T}}^{-1}, \ \ {\tilde{T}} :=
e^{\frac{\tilde{S}}{2} } e^{W}\nonumber\\
&& {\tilde{S}} := \sum_i \left ( \frac{1}{2} \frac{\partial^2}{\partial x_i^2}
+ W_i \frac{\partial}{\partial x_i} \right ) +
\sum_i W_{ii} \psi_i^{\dagger}\psi_i
+ \sum_{i \neq j} W_{ij} \psi_i^{\dagger} \psi_j .
\label{map_gen}
\eea
\noindent The
homogeneity condition on $G$ ensures that the resulting Hamiltonian has
a dynamical $OSp(2|2)$ supersymmetry with the bosonic sub-algebra
$O(2,1) \times U(1)$. The presence of the symmetry algebra
$O(2,1) \times U(1)$ in a Hamiltonian is enough to show its
equivalence to free super-oscillators. The supersymmetry of the
Hamiltonian does not play any role. The mapping to free super-oscillator
is achieved even if the $OSp(2|2)$ symmetry of a Hamiltonian is lost, but,
has only $O(2,1) \times U(1)$ symmetry. The mapping can be considered as a
necessary condition, while the construction of the complete spectrum and
associated well-behaved eigen-functions of the original Hamiltonian from the
super-oscillator basis is sufficient to claim the equivalence between
these two Hamiltonian.

The rational $BC_{N+1}$ Calogero model with $OSp(2|2)$ supersymmetry can
be mapped to free superoscillators on the half-line\cite{susy}. The complete
eigen spectra, including the degeneracy at each level, can be obtained
from the free super-oscillator basis. Relevant results in this regard are
described in section 9.4. The mapping is also applicable
to a class of $OSp(2|2)$ supersymmetric models\cite{degu} related to
short-range Dyson models\cite{nni} which are discussed in section 7. Although
infinitely many exact eigenstates can be constructed from
the superoscillator basis, the exact solvability is not known for this
class of models. Thus, an equivalence between such models with super-oscillator
can not be claimed. A careful analysis is required to check whether
the similarity operator is taking the original Hamiltonian out of its
Hilbert space or not. A similar study on the domain of the similarity operator
and the Hamiltonian is also desirable.

A comment is in order before the end of this section.
A mapping of purely rational Calogero model (i.e. $H$ without the harmonic
confinement and the fermion number operator terms) to a system of free
particles has been introduced later in Ref. \cite{gala2} by using a unitary
operator. This has been achieved at the cost of making the special conformal
generator of the underlying $O(2,1)$ symmetry non-local. An attempt to
generalize this result to the case of mapping the Calogero model with harmonic
confinement term to that of a system of free oscillators essentially leads to
non-unitary similarity operator\cite{gala2}. The non-unitary nature of the
similarity operator is in conformity with the existing
results\cite{susy}. However, the explicit form of the similarity operator
is quite different for these two cases and the readers are referred to the
relevant references\cite{susy,gala2} for details.

\section{Pseudo-hermitian Supersymmetric Systems}

The definition of a hermitian operator crucially depends on the choice of
the inner-product (/norm/metric) in the Hilbert space. The metric in the
Hilbert space is always chosen as an identity operator in the standard
treatment of quantum mechanics and a hermitian operator is defined to
be equal to its own complex-conjugate transpose. Operators not satisfying
the above criteria are termed non-hermitian and have been used extensively
to simulate dissipative quantum processes. Hermitian operators in a Hilbert
space that is endowed with an identity operator as the metric is known as
Dirac-hermitian operator in the current literature.

The question of necessity of Dirac-hermitian operators in formulating quantum
physics is as old as the subject itself. A renewed interest\cite{bend,ali,
quasi,ddt}  has been generated over the last decade in addressing the same
question in a systematic manner. The current understanding is that a quantum
system with unbroken combined Parity(${\cal{P}}$) and
Time-reversal(${\cal{T}}$) symmetry admits entirely real spectra even
though the system may be non-Dirac-hermitian. It has been
further shown that quantum system with unbroken ${\cal{PT}}$-symmetry also
admits a symmetry which is identified as a charge-conjugation (${\cal{C}}$)
symmetry. A consistent quantum description including reality of the entire
spectra and unitary time-evolution of the non-Dirac-hermitian system is
possible with the choice of a new inner-product involving the
${\cal{CPT}}$-symmetry\cite{bend}.

An alternative description of ${\cal{PT}}$-symmetric theories
is in terms of pseudo-hermitian operator\cite{ali,quasi}. An operator
$\hat{O}$ that is
related to its hermitian-conjugate ${\hat{O}}^{\dagger}$ through a
similarity transformation is defined as a pseudo-hermitian operator,
\be
{\hat{O}}^{\dagger} = \eta \hat{O} \eta^{-1}.
\ee
\noindent The hermitian conjugation of $\hat{O}$ is taken in the Hilbert space
${\cal{H}}_D$ that is endowed with the inner product
$\langle \cdot | \cdot \rangle$. The Hilbert space that is endowed with the
inner product $\langle \langle \cdot | \cdot \rangle \rangle_{\eta_+}
:= \langle \cdot |\eta_+ \cdot \rangle$ is denoted as ${\cal{H}}_{\eta_+}$.
In general, the similarity operator
$\eta$ is not unique. However, if a positive definite similarity operator
$\eta_+$ exists, the operator $\hat{O}$ can be shown to be hermitian in the
Hilbert space ${\cal{H}}_{\eta_+}$. Further, $\hat{O}$ can be mapped to a
hermitian operator $O$ through a similarity
transformation, i.e. $O = \rho \hat{O} \rho^{-1}$, where $\rho :=
\sqrt{\eta_{+}}$. Consequently, a consistent quantum description including
reality of the entire spectra and unitary time evolution is possible
for the operator $\hat{O}$ which is non-Dirac-hermitian,
but, hermitian in a Hilbert space with the metric $\eta_+$.

Several non-hermitian quantum systems admitting an entirely real spectra
and unitary time-evolution have been constructed in ${\cal{H}}_D$,
for example, in Refs. \cite{bend,ali, piju2,me1,piju4}. The ${\cal{PT}}$
symmetric extensions of Calogero models have been studied in
Refs. \cite{piju_latest,ptcsm,oldpiju,europe,smith}. A general construction of
many-particle pseudo-hermitian supersymmetric systems is presented below
which is valid for any superpotential. The pseudo-hermitian
supersymmetric rational Calogero model is included as an example. The
discussions below are based on Ref. \cite{piju_latest}.

\subsection{Isospectral Deformation}

Pseudo-hermitian quantum systems can be constructed by isospectral deformation
of known Dirac-hermitian quantum systems\cite{piju2}. The general method
involves a
realization of the basic canonical commutation relations defining the quantum
system in terms of non-Dirac-hermitian operators, which are hermitian with
respect to a pre-determined positive-definite metric in the Hilbert space.
Appropriate combinations of these operators produce a large number of
pseudo-hermitian quantum system.

\subsubsection{Deformation involving bosonic coordinates}

The metric $\eta_+^b$ in the Hilbert space ${\cal{H}}_{\eta_+^b}$
is chosen as,
\be
\eta_+^b := e^{-2 \left ( \delta \hat{B} + Re(W_-) \right ) }, \ \
W_{\pm} = \frac{1}{2} \left (W_1 \pm W_2 \right )  \ \ \delta \in R,
\label{metric}
\ee
\noindent where $W_{1,2}$ are two complex functions of the $N$ bosonic
co-ordinates. The functions $W_{+}(W_-)$ can be made real even for complex
$W_1, W_2$. For example, $W_1, W_2$ may be decomposed as,
\be
W_1= W + \chi + i \theta_1, \ \
W_2= W - \chi + i \theta_2,
\label{decomp}
\ee
\noindent where $W, \chi, \theta_1, \theta_2$ are four real functions of
the $N$ bosonic co-ordinates. A real $W_+=W$ is obtained for $\theta_1
=-\theta_2 \equiv \theta$. The operator $\hat{B}$ acts on the bosonic
co-ordinates only. The bosonic co-ordinates $x_i$ and the momenta $p_i$
are assumed to be non-hermitian in ${\cal{H}}_{\eta_+^b}$ for the type of
operator $\hat{B}$ that will be considered in this article. A set of
hermitian co-ordinates $X_i$ and momenta $P_i$ in ${\cal{H}}_{\eta_+^b}$
may be introduced as follows,
\be
X_i = \rho^{-1} x_i \rho, \ \ P_i= \rho^{-1} p_i \rho, \ \
\rho:=\sqrt{\eta_+}.
\label{inv}
\ee
\noindent The non-Dirac-hermitian operators $X_i, P_i$ trivially satisfy the
basic canonical commutation relations $\left [X_i, P_j \right ] = \delta_{ij}$.
It may be noted that the similarity transformation (\ref{inv}) keeps the
length in the momentum space as well as in the co-ordinate space invariant.

There are several choices for the operator $\hat{B}$ appearing in the metric
$\eta_+^b$ in Eq.  (\ref{metric}). For example, $\hat{B}$ may be chosen as
a linear combination of the angular momentum operators,
\be
\hat{B} := \sum_{i,j=1}^N c_{ij} \ {\cal{L}}_{ij}, \ \
{\cal{L}}_{ij}:=x_i p_j - x_j p_i,\ \ c_{ij}= -c_{ji} \in R,
\ee
\noindent with restrictions on the co-efficients $c_{ij}$ such that all the
eigenvalues of $\hat{B}$ are real. The reality condition on the eigenvalues
of $\hat{B}$ ensures the positivity of $\eta_+^b$. A simple form of $\hat{B}$
is chosen in this article:
\be
\hat{B} := {\cal{L}}_{12} = x_1 p_2 - x_2 p_1.
\ee
\noindent The co-ordinates $x_1, x_2$ and the momenta $p_1, p_2$ are not
hermitian in ${\cal{H}}_{\eta_+^b}$. It follows from the relation (\ref{inv})
that hermitian canonical conjugate operators in the Hilbert space
${\cal{H}_{\eta_+}}$ have the following expressions\cite{piju2},
\bea
&& X_1 = x_1 \ cosh \delta + i x_2 \ sinh \delta,\nonumber \\
&& X_2 = - i x_1 \ sinh \delta
+ x_2 \ cosh \delta, \ \ X_i=x_i \ for \ i > 2\nonumber \\
&& P_1 = p_1 \ cosh \delta +  i p_2 \ sinh \delta,\nonumber \\
&& P_2 = - i p_1 \ sinh \delta +
p_2 \ cosh \delta, \ P_i= p_i \ for \ i > 2.
\label{newcor}
\eea
\noindent The operator $L_{12}= X_1 P_2 - X_2 P_1={\cal{L}}_{12}$
is hermitian both in ${\cal{H}}_D$ and ${\cal{H}}_{\eta_+^b}$, ensuring
a positive-definite $\eta_+^b$.

A pseudo-hermitian Hamiltonian may be introduced as follows\cite{piju2},
\be
{\cal{H}} = \sum_{i=1}^N \Pi_i^2 + V(X), \ \
\Pi_i:= P_i + i W_{-,i}, \ \ W_{-,i} \equiv \frac{\partial W_-}{\partial X_i},
\ee
\noindent where $V(X)$ is a function of the co-ordinates $X_1, \dots,X_N$
and is hermitian in ${\cal{H}}_{\eta_+}^b$.
The real function $\theta$ appearing in $\Pi_i$ via $W_{-,i}$ can always be
rotated away by using the unitary operator $U:=e^{-i \theta}$. Without loss of
any generality, $\theta$ is chosen as zero and the generalized momentum
operators $\Pi_i$ now read,
\be
\Pi_i=P_i + i \chi_i, \ \ \chi_i \equiv \frac{\partial \chi}{\partial X_i}.
\ee
\noindent The operators $\Pi_i$ contain imaginary gauge potentials $\chi_i$.
Physical systems with imaginary gauge potentials have a wide range of
applicability\cite{piju2}. The Hamiltonian ${\cal{H}}$ is non-hermitian
in ${\cal{H}}_D$ and hermitian in ${\cal{H}}_{\eta_+^b}$.

An anti-linear ${\cal{PT}}$ transformation for the bosonic coordinates may
be introduced as follows\cite{piju2}:
\bea
&& {\cal{P}}: x_1 \leftrightarrow x_2, \ p_1 \leftrightarrow p_2, \
(x_i, p_i) \rightarrow (x_i,p_i) \ \forall \ i > 2;\nonumber \\
&& {\cal{T}}: i \rightarrow - i, \ x_i \rightarrow x_i, \ \
p_i \rightarrow - p_i;\nonumber \\
&& {\cal{PT}}: X_1 \leftrightarrow X_2, \ P_1 \leftrightarrow -P_2, \
(X_i, P_i) \rightarrow (X_i, -P_i) \ \forall i >2;\nonumber \\
&& {\cal{PT}}: \Pi_1 \leftrightarrow - \Pi_2, \ \Pi_i \rightarrow - \Pi_i \
\forall i > 2,
\label{pteqb}
\eea
\noindent where the real function $\chi$ is assumed to be invariant under
the discrete transformation ${\cal{P}}$. The operators $P_1^2(\Pi_1^2)$
or $P_2^2(\Pi_2^2)$ are not ${\cal{PT}}$-symmetric individually. However,
the combinations $P_1^2 + P_2^2$ and $\Pi_1^2 + \Pi_2^2$ are always
${\cal{PT}}$-symmetric. The Hamiltonian ${\cal{H}}$ is invariant under
${\cal{PT}}$ transformation provided the real potential $V$ remains invariant
under the transformation $X_1 \leftrightarrow X_2$.

\subsubsection{Pseudo-hermitian Realization of Clifford Algebra}

The generators (\ref{generator}) of $O(2N)$ may be used to obtain a
multi-parameter dependent pseudo-hermitian realization of the Clifford
algebra\cite{piju_latest}. The metric $\eta_+^f$ in the Hilbert space
${\cal{H}}_{\eta_+^f}$ is chosen as,
\bea
\eta_+^f & := & \prod_{i=1}^N e^{2 \gamma_i J_{i N+i}-\gamma_i} =
\prod_{i=1}^N e^{- 2 \gamma_i \psi_i^{\dagger} \psi_i}\nonumber \\
\rho^f & := & \sqrt{\eta_+^f}=\prod_{i=1}^N e^{ \gamma_i J_{i N+i}-
\frac{\gamma_i}{2}}=
\prod_{i=1}^N e^{- \gamma_i \psi_i^{\dagger} \psi_i}
\ \ \gamma_i \in R \ \forall \ i.
\label{ord}
\eea
\noindent The ordering of the generators $J_{iN+i}$ is not
required in Eq.(\ref{ord}), since the commutators $[J_{iN+i}, J_{jN+j}]=0$
for any $i$ and $j$. A set of elements $\Gamma_p$ of the real Clifford
algebra is introduced as follows,
\be
\Gamma_p  := (\rho^f)^{-1} \xi_p \rho^f,
\ee
\noindent implying the following expressions:
\bea
\Gamma_ i & = & \xi_i cosh \gamma_i
+ i \xi_{N+i} sinh \gamma_i,\nonumber \\
\Gamma_{N+i} & = & - i \xi_i sinh \gamma_i + \xi_{N+i} cosh \gamma_i,
\eea
\noindent which are hermitian in ${\cal{H}}_{\eta_+^f}$. The analog of
$\xi_{2N+1}$ in Eq. (\ref{5gamma}) is denoted as $\Gamma_{2N+1}$,
\be
\Gamma_{2N+1} := (-i)^N \Gamma_1 \Gamma_2 \dots \Gamma_{2N-1} \Gamma_{2N}=
\xi_{2N+1},
\ee
\noindent which anti-commutes with all the $\Gamma_p/\xi_p$'s and squares
to unity. The element $\Gamma_{2N+1}$ facilitates a pseudo-hermitian
realization of the generators of the group $O(2N+1)$. The readers are
referred to Ref. \cite{piju_latest} for a detail discussion on other aspects
of pseudo-hermitian realization of Clifford algebra.

The fermionic operators $\Psi_i$'s and their adjoints $\Psi_i^{\dagger}$
in ${\cal{H}}_{\eta_+^f}$,
\bea
\Psi_i & := & \frac{1}{2} \left ( \Gamma_i - i \Gamma_{N+i} \right )
= e^{-\gamma_i} \psi_i,\nonumber \\
\Psi_i^{\dagger} & := & \frac{1}{2} \left ( \Gamma_i +
i \Gamma_{N+i} \right)= e^{\gamma_i} \psi_i^{\dagger},
\eea
\noindent satisfy the basic canonical anti-commutation relations.
\be
\{\Psi_i,\Psi_j\}=0=\{\Psi_i^{\dagger},\Psi_j^{\dagger}\}, \ \
\{\Psi_i, \Psi_j^{\dagger}\}=\delta_{ij}.
\ee
\noindent The total fermion number operator $N_f$ has identical expressions,
\be
N_f=\sum_{i=1}^N \psi_i^{\dagger} \psi_i=
\sum_{i=1}^N \Psi_i^{\dagger} \Psi_i,
\ee
\noindent in ${\cal{H}}_D$ as well as in ${\cal{H}}_{\eta_+^f}$.
The relation between an eigenstate $|n_1, \dots, n_i, \dots,
n_N\rangle_{{\cal{H}}_D}$ of $N_f$ in ${\cal{H}}_D$, to the
corresponding state $|n_1, \dots, n_i, \dots,
n_N\rangle_{{\cal{H}}_{\eta_+^f}}$ in the Hilbert space ${\cal{H}}_{\eta_+^f}$
is determined as,
\be
|n_1, \dots, n_i, \dots, n_N\rangle_{{\cal{H}}_{\eta_+}} = \prod_{k=1}^N
e^{\gamma_k f_k} |n_1, \dots, n_i, \dots, n_N\rangle_{{\cal{H}}_D}, \ \
n_i = 0, 1 \ \forall \ i.
\ee
\noindent The $2^N$ states
$|n_1, \dots, n_i, \dots, n_N\rangle_{{\cal{H}}_{\eta_+^f}}$
form a complete set of orthonormal states in ${\cal{H}}_{\eta_+^f}$,
while $|n_1, \dots, n_i, \dots, n_N\rangle_{{\cal{H}}_{D}}$ constitute a
complete set of orthonormal states in ${\cal{H}}_D$. The action of
$\Psi_i(\Psi_i^{\dagger})$ on
$|n_1, \dots, n_i, \dots, n_N\rangle_{{\cal{H}}_{\eta_+^f}}$
is identical to that of $\psi_i(\psi_i^{\dagger})$ on
$|n_1, \dots, n_i, \dots, n_N\rangle_{{\cal{H}}_D}$:
\bea
\Psi_i |n_1, \dots, n_i, \dots n_N\rangle_{{\cal{H}}_{\eta_+}} & = & 0,
\ \ \ \
if \ \ \ \ n_i=0,\nonumber \\
& = & |n_1, \dots, 0, \dots n_N\rangle_{{\cal{H}}_{\eta_+}}, \ \ \ \
if \ \ \ \ n_i=1,\nonumber \\
\Psi_i^{\dagger} |n_1, \dots, n_i, \dots n_N\rangle_{{\cal{H}}_{\eta_+}}
& = & 0, \ \ \ \ if \ \ \ \ n_i=1,\nonumber \\
& = & |n_1, \dots, 1, \dots n_N\rangle_{{\cal{H}}_{\eta_+}}, \ \ \ \
if \ \ \ \ n_i=0.
\eea
\noindent The pseudo-hermitian odd elements
$\Gamma_i$ or $\Psi_i, \Psi_i^{\dagger}$ is used to
construct pseudo-hermitian supersymmetric quantum systems.
The fermionic permutation operator $\tilde{K_{ij}}$,
\be
\tilde{K}_{ij} := \frac{1}{2} \left [ \Psi_i^{\dagger} - \Psi_j^{\dagger},
\Psi_i - \Psi_j \right ] = 1 - \left ( \Psi_i - \Psi_j \right ) \left (
\Psi_i^{\dagger} - \Psi_j^{\dagger} \right ) = K_{ij},
\ee
\noindent is hermitian in ${\cal{H}}_D$ as well as in ${\cal{H}}_{\eta_+}$.
The metric $\eta_+^f$ is invariant under the action of $\tilde{K}_{ij}$.
As in the case of bosonic coordinates in ${\cal{H}}_{\eta_+^b}$, an anti-linear
${\cal{PT}}$ transformation for the elements $\xi_p$ may be introduced
as follows:
\bea
{\cal{T}}: && i \rightarrow -i, \ \xi_p \rightarrow \xi_p \ \forall
\ p;\nonumber \\
{\cal{P}}: && \xi_i \rightarrow \tilde{\xi}_i = \xi_i cos \beta +
\xi_{N+i} sin \beta,\nonumber \\
&& \xi_{N+i} \rightarrow \tilde{\xi}_{N+i} = \xi_i sin \beta -
\xi_{N+i} cos \beta, \ \ 0 \leq \beta \leq 2 \pi,
\eea
\noindent where $\beta$ appears as a phase which may be fixed at
some specific value depending on the physical requirements.
The action of the ${\cal{PT}}$ transformation on the
fermionic variables $\psi_i$ is as follows,
\bea
&& {\cal{P}}: \psi_i \rightarrow e^{-i \beta} \psi_i^{\dagger}, \ \
\psi_i^{\dagger} \rightarrow e^{ i \beta} \psi_i,\nonumber \\
&& {\cal{T}}: \psi_i \rightarrow \psi_i^{\dagger}, \ \
\psi_i^{\dagger} \rightarrow \psi_i\nonumber \\
&& {\cal{PT}}: \psi_i \rightarrow e^{i \beta} \psi_i, \ \
\psi_i^{\dagger} \rightarrow e^{-i \beta} \psi_i^{\dagger}. \
\label{pteqf}
\eea
\noindent The supersymmetric Hamiltonian in Eq. (\ref{eq4}) contains bi-linear
terms of the form $\psi_i^{\dagger} \psi_j$ which are ${\cal{PT}}$-invariant
for any $\beta$. The Hamiltonian (\ref{h2}) in the `atypical' representation
of $OSp(2|2)$ contains bi-linear terms of the form $\psi_i \psi_j,
\psi_i^{\dagger} \psi_j^{\dagger}$ which are ${\cal{PT}}$-invariant only
for $\beta=0$. However, $\Gamma_{2N+1}$ appearing in the same Hamiltonian
is ${\cal{PT}}$-invariant for any $\beta$. This can be checked by expressing
$\Gamma_{2N+1}$  in terms of fermionic variables as,
\be
\Gamma_{2N+1} = (-1)^N \prod_{i=1}^N \left ( 2 \psi_i^{\dagger} \psi_i
-1 \right ).
\ee
\noindent Similarly, the fermionic exchange operator $\tilde{K}_{ij}=K_{ij}$
is invariant under ${\cal{PT}}$ for any $\beta$. This provides a framework for
constructing ${\cal{PT}}$ symmetric supersymmetric non-Dirac-hermitian
Hamiltonian.

\subsection{Many-particle non-Dirac-hermitian Supersymmetry}

The metric $\eta_+$ in the Hilbert space ${\cal{H}}_{\eta_+}:=
{\cal{H}}_{\eta_+^b} \otimes {\cal{H}}_{\eta_+^f}$ of the supersymmetric
Hamiltonian is chosen as,
\be
\eta_+:= \eta_+^b \otimes \eta_+^f, 
\label{metric_tot}
\ee
\noindent where $\eta_+^b$ and $\eta_+^f$ are given by Eqs. (\ref{metric}) and
(\ref{ord}), respectively. The supercharges are introduced as follows,
\bea
&& \tilde{Q}_1 = \sum_{i=1}^N e^{-\gamma_i} \ \psi_i  \left ( P_i +
i W_{1,i} \right )\nonumber \\
&& \tilde{Q}_2 = \sum_{i=1}^N e^{\gamma_i} \ \psi_i^{\dagger} \left ( P_i -
i W_{2,i} \right ),\nonumber \\
&& W_{1,i} = \frac{\partial W_{1}} {\partial X_i}, \ \
\ \ W_{2,i} = \frac{\partial W_{2}} {\partial X_i}, 
\eea
\noindent leading to the the supersymmetric Hamiltonian,
\bea
\tilde{H} & := & \frac{1}{2} \{\tilde{Q}_1, \tilde{Q}_2\}\nonumber \\
& = & \frac{1}{2} \sum_{i=1}^N \left [ \Pi_i^2 + \left ( W_{+,i} \right )^2 -
W_{+,ii} \right ] + \sum_{i,j=1}^N e^{\gamma_i-\gamma_j} W_{+,ij}
\psi_i^\dagger \psi_j,\nonumber \\
W_{\pm,i} & \equiv & \frac{1}{2} \left ( W_{1,i} \pm W_{2,i} \right ), \ \
W_{+,ij} \equiv \frac{\partial^2 W_+}{\partial X_i \partial X_j}. \
\eea
\noindent The last term in $\tilde{H}$ is manifestly non-hermitian in
${\cal{H}}_D$. The non-hermiticity of the generalized momentum operators
in ${\cal{H}}_D$ appears due to the presence of imaginary gauge potentials
$\chi_i$.  It is worth re-emphasizing that imaginary gauge potentials appear in
the study of diverse branches of physics including metal-insulator
transitions or depinning of flux lines from extended defects in type-II
superconductors and unzipping of DNA\cite{piju2}. The purely bosonic potentials
containing in the second and the third terms are functions of the coordinates
$X_i$ and are in general non-hermitian in ${\cal{H}}_D$. The Hamiltonian
$\tilde{H}$ is hermitian in
${\cal{H}}_{\eta_+}$, provided the complex functions $W_{1,2}$ are taken
to be of the form  (\ref{decomp}) with $\theta_1= -\theta_2 \equiv \theta$.
Finally, $\theta$ can gauged away from the Hamiltonian through a unitary
transformation. The hermiticity of $\tilde{H}$ in ${\cal{H}}_{\eta_+}$
may be checked by re-expressing it as,
\be
\tilde{H} = \frac{1}{2} \sum_{i=1}^N \left [ \Pi_i^2
+ \left ( W_{i} \right )^2 - W_{ii} \right ] +
\sum_{i,j=1}^N W_{ij} \Psi_i^\dagger \Psi_j.
\ee
\noindent It may be noted that the non-Dirac-hermitian operators $\Pi_i=
p_i + i \chi_i $ and non-Dirac-hermitian functions $W_i, W_{ij}$
are hermitian in ${\cal{H}}_{\eta_+}$. The Hamiltonian $\tilde{H}$ is
isospectral with the Dirac-hermitian Hamiltonian $H$ in Eq. (\ref{eq4}),
\be
H = \left ( U \rho \right ) \tilde{H} \left ( U \rho^{-1} \right ).
\ee
\noindent An exactly solved non-Dirac-hermitian quantum system $\tilde{H}$
may be constructed for the choice of $W$ for which exactly solvable
many-particle supersymmetric quantum systems $H$ is known.
A set of orthonormal eigenfunctions $\chi_n$ of $\tilde{H}$ in
${\cal{H}}_{\eta_+}$ may be constructed from the orthonormal eigenfunctions
$\Phi_n$ of $H$ in $H_D$ by using the relation, $\chi_n = (U\rho)^{-1} \Phi_n$.

The main focus of this article is on systems with inverse-square interactions.
The superpotential $W$ is chosen as,
\be
W(X_1, X_2, \dots, X_N) \equiv W_0 = - ln G(X_1, X_2, \dots, X_N)
\ee
\noindent where $G$ is a homogeneous function of degree
$d$,
\be
\sum_{i=1}^N X_i \frac{\partial G(X_1, \dots, X_N)}{\partial X_i} =
d G(X_1, \dots, X_N).
\ee
\noindent The Hamiltonian $\tilde{H}_0$ corresponding to $W_0$, along with
the dilatation operator $\tilde{D}$ and the conformal operator $\tilde{K}$,
\be
\tilde{D} = -\frac{1}{4} \sum_{i=1}^N \left ( X_i \Pi_i + \Pi_i X_i \right ),
\ \ \tilde{K} = \frac{1}{2} \sum_{i=1}^N X_i^2= K,
\ee
\noindent satisfy the $O(2,1)$ algebra which appears as a bosonic sub-algebra
of the $OSp(2|2)$ super-group. The `typical' representation of the $OSp(2|2)$ is
realized with the following definition of the operators:
\be
\tilde{S}= \frac{1}{{2}} \sum_{i=1}^N e^{-\gamma_i} \psi_i X_i, \ \
\tilde{S}^{\dagger} = \frac{1}{{2}} \sum_{i=1}^N e^{\gamma_i} \psi_i^{\dagger}
X_i, \ \
\tilde{Y} = Y.
\ee
\noindent The generators $\tilde{H}_0$, $\tilde{D}$,
$\tilde{Q}_1$, $\tilde{Q}_2$, $\tilde{S}$ and $\tilde{S}^{\dagger}$ of
$OSp(2|2)$ are hermitian
in ${\cal{H}}_{\eta_+}$. The Dirac-hermitian generators $K$ and $Y$ are also
hermitian in ${\cal{H}}_{\eta_+}$. Similarly, the `atypical' representation of
$OSp(2|2)$ in ${\cal{H}}_{\eta_+}$ is obtained trivially by replacing
$x_i \rightarrow X_i, p_i \rightarrow \Pi_i$ and
$\xi_p \rightarrow \Gamma_p$ in the corresponding expressions of the
generators in ${\cal{H}}_{D}$. It may be recalled in this regard that
${\cal{L}}_{12}=L_{12}$ and $\xi_{2N+1} = \Gamma_{2N+1}$.

\subsection{Examples: Rational Calogero Models}

The pseudo-hermitian supersymmetric rational Calogero model is presented as
an example for which the superpotential is chosen as,
\be
W = - \lambda \sum_{i<j=1}^N ln \left ( X_i - X_j \right )
+ \frac{1}{2} \sum_{i=1}^N X_i^2,
\ee
\noindent and the corresponding non-Dirac-hermitian Hamiltonian reads,
\bea
\tilde{H} & = & \frac{1}{2} \sum_{i=1}^N \Pi_i^2
+ \frac{1}{2} \lambda (\lambda-1) \sum_{i \neq j=1}^N X_{ij}^{-2} +
\frac{1}{2}\sum_{i=1}^N  x_i^2\nonumber \\
& + & \lambda \sum_{i \neq j=1}^N X_{ij}^{-2} \left ( \psi_i^{\dagger} \psi_i
- e^{\gamma_i-\gamma_j} \psi_i^{\dagger} \psi_j \right )+
\sum_{i=1}^N \psi_i^{\dagger} \psi_i
-\frac{N}{2} - \frac{\lambda}{2} N (N-1),\nonumber \\
X_{12} & = & \left ( x_1 - x_2 \right ) cosh \delta + i \left ( x_1 +
x_2 \right ) sinh \delta,\nonumber \\
X_{1j} & = &  x_1 cosh \delta + i x_2 sinh \delta - x_j, \ \ j > 2,\nonumber \\
X_{2 j} & = &  -i x_1 sinh \delta + x_2 cosh \delta - x_j \ \ j > 2,\nonumber \\
X_{ij} & = & x_i - x_j, \ \  (i, j) > 2.
\label{calo}
\eea
\noindent The following differences between the system governed by $\tilde{H}$
and the standard rational Calogero Hamiltonian in Eq. (\ref{eq6.2}) are to be
noted. The many-body inverse-square interaction
term in $\tilde{H}$ is neither invariant under translation nor singular
for $x_{1} = x_i, i > 1$ and  $x_2 = x_i, i > 2$. The permutation symmetry
(\ref{permu}) of $H$ is no longer a symmetry of $\tilde{H}$.
However, the Hamiltonian $\tilde{H}$ is invariant under a combined
${\cal{PT}}$ operation as defined in Eqs. (\ref{pteqf}) and (\ref{pteqb}).
The Hamiltonian $\tilde{H}$ and the Dirac-hermitian rational $A_{N+1}$ Calogero
model are related to each other through a non-unitary similarity transformation.
This implies that these models are isospectral provided identical boundary
conditions have been used. However, there are no compelling reasons to solve
these systems under identical boundary conditions. The Hamiltonian $\tilde{H}$
has $(N-3)(N-2)$ number of less singular points compared to the standard
Calogero model\cite{cs} due to the non-singular points at $x_{1} = x_i,
i > 1$ and  $x_2 = x_i, i > 2$. Thus, the configuration spaces of these two
Hamiltonians are different allowing different boundary conditions. It is to be
seen whether a consistent description of ${\tilde{H}}$, including an entirely
real spectrum,  is allowed for any modified boundary condition or not.

\section{Quantum Systems in Higher Dimensions}

There are many higher dimensional generalizations of the rational Calogero
model for which infinitely many exact eigen states and eigen values can be
obtained analytically\cite{marchio,mbs,mri}. The exact eigenstates owe their
existence to mainly the underlying $O(2,1)$ symmetry. However, not a single
model belonging to this class is exactly solved for a complete set of states. 
Nevertheless, the study of theses systems gives a better understanding of
quantum systems in higher dimensions. The Calogero-Marchioro model\cite{marchio}
is one such example which describes a $D >1$ dimensional quantum
system of particles interacting with each other through two-body and three-body
inverse-square interaction terms. The importance of $D=2$ dimensional
Calogero-Marchioro model lies in its relevance in the study of a host of
different subjects, like normal matrix model\cite{kray,rsu,go,joshua}, two
dimensional Bose system\cite{rsu}, quantum Hall effect\cite{kklz}, quantum
dot\cite{imsc}, extended superconformal symmetry\cite{pkg}. It is known that
the two dimensional Calogero-Marchioro model at some specific value of
the coupling constant describes the dynamics of a Gaussian ensemble of normal
matrices in the large $N$ limit\cite{kray,rus,go}. The low energy limit of
$2+1$ dimensional Yang-Mills theory, dimensionally reduced to $0+1$ dimensions,
is described by the Gaussian action of normal matrices\cite{pkg}. Thus, the
Calogero-Marchioro model is also indirectly related to $2+1$ dimensional
Yang-Mills theory. The supersymmetric version of the $D$ dimensional
Calogero-Marchioro model is presented in this section.

It has been suggested\cite{gib} that the rational Calogero model
with extended ${\cal{N}}=4$ $SU(1,1|2)$ superconformal symmetry may describe
the motion of a test super-particle in the near-horizon geometry of $3+1$
dimensional extremal black holes. The first initiative to construct such a
model was not completely successful\cite{nw}. In particular, the
resulting Hamiltonian is $SU(1,1|2)$ superconformal only for spercific values
of the strength of the inverse-square interaction. Several attempts have been
made thereafter to construct models with ${\cal{N}}=4$ superconformal 
symmetry\cite{sc,belu,olaf2,olaf3,gala1,gala3,gala4,gala5,kiri}. The
development in this
regard has been described in the review article \cite{review_scm}, which also
contains a few examples of generalized Calogero-type models in diverse
dimensions. An example of ${\cal{N}}=4$ superconformal model has been
constructed in Ref. \cite{kiri}, whose bosonic sector is not described
by the standard rational Calogero model. In fact, the construction of a
${\cal{N}}=4$ superconformal Hamiltonian for arbitrary number of particles
and the generic values of the coupling constant, which reduces to the standard
Calogero or Calogero-Marchioro model in the purely bosonic sector, is still
beyond the reach in $D \neq 2$ dimensions. It is shown that the $D=2$
dimensional Calogero-Marchioro model can naturally be embedded into an
extended ${\cal{N}}=4$ $SU(1,1|2)$ superconformal Hamiltonian\cite{pkg}.
The construction of rational Calogero-Marchioro model with $SU(1,1|2)$
superconformal symmetry is discussed in some detail. 

\subsection{Supersymmetric Calogero-Marchioro Model}

The supercharge ${\cal{Q}}_1$ and its conjugate ${\cal{Q}}_1^{\dagger}$ are
defined as,
\bea
&& {\cal{Q}}_1=\sum_{i,\mu} \psi_{i,\mu}^{\dagger} \ {\cal{A}}_{i,\mu}, \ \ \ \
{\cal{Q}}_1^\dagger = \sum_{i,\mu} \psi_{i,\mu} \  
{\cal{A}}_{i,\mu}^{\dagger},\nonumber\\
&& {\cal{A}}_{i,\mu} := p_{i,\mu} - i W_{i\mu},
\ \ {\cal{A}}_{i,\mu}^{\dagger}:= p_{i,\mu} +
i W_{i,\mu},\nonumber \\
&& p_{i,\mu} = - i \frac{\partial}{\partial x_{i,\mu}}, \ \
W_{i,\mu} = \frac{\partial W}{\partial x_{i,\mu}}, \ \ i=1, \dots, N, \
\mu,\nu= 1, \dots, D,
\label{eq6}
\eea
\noindent where $W$ is the superpotential and the $N D$ fermionic variables
$\psi_{i,\mu}$'s satisfy the Clifford algebra,
\be
\{\psi_{i,\mu},\psi_{j,\nu}\}=0=\{\psi_{i,\mu}^{\dagger},\psi_{j,\nu}
^{\dagger}\}, \ \
\{\psi_{i,\mu}, \psi_{j,\nu}^{\dagger}\}=\delta_{ij} \delta_{\mu,\nu}.
\label{eq7}
\ee
\noindent The superpotential is chosen as,
\be
W \equiv W_0 = - g \sum_{i <j} ln {\mid \vec{r}_{ij} \mid},
\ \ \vec{r}_{ij} \equiv \vec{r}_i - \vec{r}_j,
\label{eq8.1}
\ee
\noindent which results in the following supersymmetric Hamiltonian,
\bea
H_0 & = & \frac{1}{2} \sum_{i,\mu } p_{i,\mu }^2 + \frac{g}{2} ( g + D-2)
\sum_{i \neq j} \vec{r}_{ij}^{-2} + \frac{g^2}{2}  \sum_{i \neq j \neq k}
\left ( \vec{r}_{ij} . \vec{r}_{ik} \right ) \vec{r}_{ij}^{- 2}
\vec{r}_{ik}^{-2}\nonumber \\
& + & g \sum_{i \neq j; \mu} \left ( 2 \left ( x_{i,\mu} -
x_{j,\mu} \right )^2 \vec{r}_{ij}^{-2} - 1 \right )
\vec{r}_{ij}^{-2} \left (\psi_{i,\mu}^{\dagger} \psi_{i,\mu} -
\psi_{i,\mu}^{\dagger} \psi_{j,\mu} \right )\nonumber \\
& + & 2 g \sum_{i \neq j; \mu \neq \nu} \left ( x_{i,\mu} -
x_{j,\mu} \right )
\left ( x_{i,\nu} - x_{j,\nu} \right ) \vec{r}_{ij}^{-4} \left (
\psi_{i,\mu}^{\dagger} \psi_{i,\nu} -
\psi_{i,\mu}^{\dagger} \psi_{j,\nu} \right ).
\label{eq10}
\eea
\noindent The Hamiltonian $H_0$ along with the Dilatation operator $D$ and
the conformal generator $K$,
\be
{D} = -\frac{1}{4} \sum_{i,\mu } \{ x_{i,\mu }, p_{i,\mu } \}, \ \
K = \frac{1}{2} \sum_{i,\mu } x_{i,\mu }^2, \ \
\ee
\noindent satisfy the $O(2,1)$ algebra given in Eq. (\ref{ap_II.2}).

The super-Hamiltonian $H_0$ does not have a normalizable
ground-state. The quantum evolution of the system can instead be studied
in terms of the operator $R$ or $H$: 
\be
H =  R + B - T, \ \ R = H_0 + K, \ \
B = \frac{1}{2} \sum_{i,\mu} \left [\psi_{i,\mu}^{\dagger},
\psi_{i,\mu} \right ], \ \ T=\frac{g}{2} N (N-1) .
\label{eq11}
\ee
\noindent The zero-fermion sector of the supersymmetric Hamiltonian $H$
describes $D$ dimensional Calogero-Marchioro Hamiltonian. The supersymmetric
rational $A_{N+1}$ Calogero model is obtained from (\ref{eq11}) for $D=1$.
The supersymmetric ground state of $H$ is obtained in the region $g > 0$,
\be
\Phi_0= \phi_0 |0\rangle, \ \ \phi_0 \equiv
\prod_{i<J} {\mid r_{ij} \mid}^g e^{-\frac{1}{2} \sum_i \vec{r}_i^2},
\ee
\noindent where $|0\rangle$ is now the fermionic vacuum in $2^{ND}$ dimensional
Fock space. A set of exact eigenstates is constructed below in the
supersymmetric phase. The analysis in the supersymmetry-breaking phase($g < 0$)
is similar to the case of one dimensional supersymmetric rational Calogero
model and is given in Ref. \cite{pkg}.

The complete $OSp(2|2)$ algebra is realized by the introduction of the
following operators,
\bea
{\cal{S}}_1 & := & \sum_{i,\mu} \psi_{i,\mu}^{\dagger} x_{i,\mu},\ \
{\cal{S}}_1^{\dagger} := \sum_{i,\mu} \psi_{i,\mu} x_{i,\mu}, \nonumber \\
{\cal{F}}_1 & = & {\cal{Q}}_1 - i {\cal{S}}_1, \ \
{\cal{F}}_2 = {\cal{Q}}_1^{\dagger}- i {\cal{S}}_1^{\dagger},\nonumber \\
{\cal{F}}_1^{\dagger} & = & {\cal{Q}}_1^{\dagger} + i {\cal{S}}_1^{\dagger}, \ \
{\cal{F}}_2^{\dagger}= {\cal{Q}}_1 + i {\cal{S}}_1,\nonumber \\
\label{eq13}
\eea
\noindent The supersymmetric Hamiltonian $H$ in Eq. (\ref{eq11}) is
re-expressed in terms
of the operators ${\cal{F}}_1, {\cal{F}}_1^{\dagger}$ as,
$ H = \frac{1}{2} \{{\cal{F}}_1, {\cal{F}}_1^{\dagger} \}$.
\noindent The following algebra,
\bea
&& {\cal{B}}_2^{\dagger} := - \frac{1}{4} \{ {\cal{F}}_1^{\dagger},
{\cal{F}}_2^{\dagger} \},
\nonumber \\
&& [H, {\cal{B}}_2^{\dagger}] = 2 {\cal{B}}_2^{\dagger}, \ \
[H, {\cal{F}}_2^{\dagger}] = 2 {\cal{F}}_2^{\dagger},
\eea
\noindent allows a construction of the excited states,
\be
\Phi_{n,\nu}= {\cal{B}}_2^{{\dagger}^n} {\cal{F}}_2^{{\dagger}^{\nu}}
\Phi_0,
\ee
\noindent with the energy eigen values $E_{n,\nu}= 2 ( n + \nu ) $. The bosonic
quantum number $n$ can take any non-negative integer values, while the
fermionic quantum number $\nu=0, 1$. The spectrum does not reduce to that
of $D$ dimensional $N$ free superoscillators in the limit $g \rightarrow 0$.
The set of exact eigenstates (\ref{eq16}) is thus incomplete and the
complete spectrum is not known.

\subsection{Extended Superconformal Symmetry}

The coefficients of the bosonic two-body and the thee-body interaction terms
are identical for $D=2$. Consequently, the supersymmetric Hamiltonian $H$
can be embedded into an extended ${\cal{N}}=4$ superconformal symmetry
in $D=2$ space dimensions. The general form of the superpotential in $D=2$
that gives rise to Hamiltonian with extended superconformal symmetry may
be expressed as\cite{pkg},
\bea
&& W_0= - ln G, \ \
G  =  f(z_1, z_2, \dots, z_N) \ g(z_1^*, z_2^*, \dots, z_N^*),\nonumber \\
&& \ \ z_k = x_{k,1} + i x_{k,2}, \ \
\ \ z_k^* = x_{k,1} - i x_{k,2}, \ \
\eea
\noindent where $G$ is a homogeneous function. The homogeneity condition
on $G$ implies that the (anti-)holomorphic function $(g)f$ should also be
homogeneous. The superpotentials of the Calogero-Marchioro model and a
nearest-neighbor variant of this model\cite{nni} in $D=2$ satisfy the above
criteria. The rest of the discussions is based on two dimensional
superconformal Calogero-Marchioro model\cite{pkg}.

The Hamiltonian $H_0$ has an internal $SU(2)$ symmetry. The generators
of the $SU(2)$ are the even operator $B$ in Eq. (\ref{eq11}) and the
operators,
\be
Y = \frac{1}{2} \sum_i \epsilon_{\mu \nu} \psi_{i,\mu}
\psi_{i,\nu}, \ \
Y^{\dagger} = - \frac{1}{2} \sum_i \epsilon_{\mu \nu}
\psi_{i,\mu}^{\dagger} \psi_{i,\nu}^{\dagger}, \
\label{eq19.1},
\ee
\noindent satisfying the algebra
\be
[Y, Y^{\dagger}] = - B, \ \ [B, Y] = - 2 Y, \ \ [ B, Y^{\dagger}] =
2 Y^{\dagger}.
\ee
\noindent The repeated indices of the Levi-Civita pseudo-tensor
$\epsilon_{\mu \nu}$ are always summed over. The $SU(1,1,|2)$ requires
the introduction of the following odd operators:
\bea
{\cal{Q}}_2  & = & \sum_i \epsilon_{\mu,\nu} \psi_{i,\nu} {\cal{A}}_{i, \mu},
\ \
{\cal{Q}}_2^{\dagger} = \sum_i \epsilon_{\mu,\nu} \psi_{i,\nu}^{\dagger}
{\cal{A}}_{i, \mu}^{\dagger}\nonumber \\
{\cal{S}}_2 & = & \sum_i \epsilon_{\mu \nu} \psi_{i, \nu} x_{i,\mu},\ \
{\cal{S}}_2^{\dagger} = 
\sum_i \epsilon_{\mu \nu} \psi_{i, \nu}^{\dagger} x_{i,\mu}.
\eea
\noindent Defining two new odd operators in terms of ${\cal{Q}}_2, {\cal{S}}_2$
and their adjoints,
\be
{\tilde{\cal{F}}}_1:={\cal{Q}}_2 - i {\cal{S}}_2, \ \
{\tilde{\cal{F}}}_2:={\cal{Q}}_2^{\dagger} - i {\cal{S}}_2^{\dagger}, \ \
\ee
\noindent the following algebra holds,
\bea
&& \frac{1}{2} \{ \tilde{\cal{F}}_2, \tilde{\cal{F}}_2^{\dagger} \}
= H + 2 T, \ \
\frac{1}{2} \{ {\cal{F}}_1, \tilde{\cal{F}}_2^{\dagger} \} = - \frac{1}{2}
\{ {\cal{F}}_1^{\dagger}, \tilde{\cal{F}}_2 \} = - i {\tilde{J}},\ \
\frac{1}{2} \{ {\cal{F}}_2, {\cal{F}}_2^{\dagger} \} = {\tilde{H}},\nonumber \\
&& \frac{1}{2} \{ \tilde{\cal{F}}_1, \tilde{\cal{F}}_1^{\dagger} \} =
\tilde{H} - 2 T,\ \
\frac{1}{2} \{ {\cal{F}}_2, \tilde{\cal{F}}_1^{\dagger} \} = - \frac{1}{2}
\{ {\cal{F}}_2^{\dagger}, \tilde{\cal{F}}_1 \} = - i \tilde{J}, \ \
\tilde{H}:=H - B + 2 T\nonumber \\.
\label{non-diag}
\eea
\noindent The operator angular momentum operator $\tilde{J}$ appearing above
is defined as,
\be
\tilde{J} = \sum_i \epsilon_{\mu \nu} \left ( x_{i,\nu} p_{i, \mu} + i
\psi_{i, \mu}^{\dagger} \psi_{i,\nu} \right ).
\ee
\noindent All other non-vanishing anti-commutators are,
\bea
&& - \frac{1}{2} \{ {\cal{F}}_1, \tilde{\cal{F}}_1^{\dagger} \} = \frac{1}{2}
\{ \tilde{\cal{F}}_2, {\cal{F}}_2^{\dagger} \} = 2 Y^{\dagger},\nonumber \\
&& - \frac{1}{2} \{ \tilde{\cal{F}}_1, {\cal{F}}_1^{\dagger} \} = \frac{1}{2}
\{ {\cal{F}}_2, \tilde{\cal{F}}_2^{\dagger} \} = 2 Y,\nonumber \\
&& \frac{1}{4} \{ {\cal{F}}_1, {\cal{F}}_2 \} =
\frac{1}{4} \{ \tilde{\cal{F}}_1, \tilde{\cal{F}}_2 \}
= - {\cal{B}}_2, \ \
\eea
\noindent The evolution can be described either in terms of the Hamiltonian
$H$ or its dual $\tilde{H}$. The dual Hamiltonian $\tilde{H}$ is used to
study the spectrum in supersymmetry-breaking phase $g <0$\cite{pkg}. The
algebra in Eq. (\ref{non-diag}) is not in a diagonal form because of the
presence of $\tilde{J}$. However, a diagonal form of the algebra can be
obtained\cite{pkg} by introducing two new supercharges as a linear combination
of ${\cal{F}}_1$ and $ \tilde{\cal{F}}_2$.

\section{Systems with Internal Degrees of Freedom}

The Calogero models with internal degrees of freedom have been studied
previously in the literature\cite{poly}. Such models naturally appear
in the reductions of various matrix models to many-particle quantum systems.
A supersymmetric Calogero-type model with internal $U(2)$ degrees of freedom
has been obtained via the reductions of certain gauged matrix
models\cite{olaf1}. The relevant discussions in this regard have been included
in the review article\cite{review_scm}. Calogero models with $k$ numbers of
internal degrees of freedom may also be introduced directly in terms of
permutation operators constructed out of the generators of
$SU(k)$\cite{mina,poly}.
A generalization of the Calogero-Sutherland models, where the indices
corresponding to the particles internal degrees of freedom form a
representation of the $gl(n|m)$ graded lie algebra, has also been considered
in the literature\cite{hs1,ahn,biru2,biru4}. These models are integrable
and have many interesting properties. These models are also termed 
`supersymmetric'\cite{hs1}, because of the presence of the `graded permutation
operator' in the Hamiltonian. However, the full superalgebra (\ref{salgebra})
is realized for the corresponding quantum Hamiltonians only for the simplest
case of $gl(1|1)$. Thus, discussions of this class of models with $n \neq 1,
m \neq 1$ are beyond the scope of this paper.

The supersymmetric Hamiltonian $H$ in Eq. (\ref{eq4}) has a block-diagonal
structure in the
fermionic representation, thereby, giving internal structures to the bosonic
particles, which may not be always interpreted in terms of spin degrees of
freedom. It may be recalled that the fermionic exchange operator appearing
in the supersymmetric rational Calogero Hamiltonian realizes a tensor
representation of symmetric group $S_N$ of fermionic operators\cite{rus}.
A class of supersymmetric many-particle Hamiltonians is constructed in this
section which can be interpreted as $N$ bosonic particles with internal spin
degrees of freedom. 

Several spin chain Hamiltonians, including the celebrated
Haldane-Shastry model\cite{hs,hs1}, may be obtained from Calogero models with
internal degrees of freedom in the strong interaction limit, known as
`freezing limit'\cite{poly}. The supersymmetric Hamiltonians considered in
this section reduce to $XY$ model on a non-uniform lattice in the `freezing
limit'. The `freezing limit' for a supersymmetric system may be taken in the
following way. The supersymmetric Hamiltonian (\ref{eq4}) can be re-written as,
\be
H = \frac{1}{2} \sum_i \left ( p_i^2 + W_i^2 \right ) + \frac{1}{2}
\sum_{i,j} W_{ij} \left ( \psi_i^{\dagger} \psi_j - \psi_i \psi_j^{\dagger}
\right ).
\label{rbm_hami}
\ee
\noindent The bosonic and the fermionic parts of the supersymmetric Hamiltonian
decouple from each other for a superpotential satisfying $W_{ij} = constant \
\forall \ i,j$. This implies that the superpotential is a quadratic form of the
bosonic co-ordinates. The Hamiltonian of superoscillators is one such example.
In general, the bosonic and the fermionic degrees of freedom can not be
decoupled for any other choices of $W$. However,
The fermionic part can be decoupled from the parent Hamiltonian in the `freezing
limit'\cite{poly}. In particular, the superpotential $W$ is taken to be
proportional to an overall coupling constant $\lambda$. The coefficient of the
bosonic potential term $W_i^2$ becomes $\lambda^2$, while it is $\lambda$ for
the fermionic part of the Hamiltonian. An effective Hamiltonian in the
strong interaction limit $\lambda \rightarrow \infty$ may be obtained
by first multiplying $H$ with $\lambda^{-2}$ and then taking
$\lambda \rightarrow \infty$. The leading relevant term in this limit is,
\be
H \equiv \frac{1}{2 \lambda}
\sum_{i,j} W_{ij} \left ( \psi_i^{\dagger} \psi_j - \psi_i \psi_j^{\dagger}
\right ) + O \left ( \lambda^{-2} \right ),
\label{eq6.11}
\ee
\noindent where the bosonic coordinates in $W_{ij}$ take the value of
their classical minimum equilibrium configurations, $W_i=0$.
The non-dynamical term $\sum_i W_i^2$ vanishes identically for the classical
minimum equilibrium configurations. The Hamiltonian $H$ in Eq. (\ref{eq6.11})
contains only bilinear terms in the fermionic operators and is always
diagonalizable.

\subsection{Supersymmetry \& XY model}

The superpotential $W$ is chosen such that,
\bea
W_{ij} & = & \delta_{ij} g_i (x_1, x_2, \dots, x_N) + \delta_{i,j+1}
h_{i-1} (x_1, x_2, \dots, x_N)\nonumber \\
& + & \delta_{i,j-1} h_i (x_1, x_2, \dots, x_N),
\label{sp_nn}
\eea
\noindent where $h_i$'s and $g_i$'s are arbitrary functions of the bosonic
coordinates. The supersymmetric Hamiltonian (\ref{eq4}) now reads,
\be
H = \frac{1}{2} \sum_i \left ( p_i^2 + W_i^2 \right ) +
\frac{1}{2} \sum_i \left [ g_i ( 2 n_i - 1) 
+ 2 h_i \left ( \psi_i^{\dagger} \psi_{i+1}
- \psi_i \psi_{i+1}^{\dagger} \right ) \right ].
\label{eq6.3}
\ee
\noindent The Hamiltonian $H$ can be re-written in terms of spin degrees
freedom by using the Jordan-Wigner transformation(See Appendix V) with
the periodic boundary conditions:
\be
H = \frac{1}{2} \sum_i \left ( p_i^2 + W_i^2 \right ) +
\frac{1}{2} \sum_i \left [ g_i \sigma_i^z + h_i \left (
\sigma_i^{x} \sigma_{i+1}^x + \sigma_i^y \sigma_{i+1}^y \right ) \right ].
\label{eq6.33}
\ee
\noindent The use of the Jordan-Wigner transformation and hence,
the interpretation of $H$ as a system of spin-$\frac{1}{2}$ particles
interacting with each other through inverse-square interactions fails,
if the fermion-fermion interaction is beyond the next-neighbour.
The situation is saved for the particular choice of the superpotential in
Eq. (\ref{sp_nn}). The supersymmetric Hamiltonian $H$, in general,
describes an $N$ particle system with both kinematic and internal
spin degrees of freedom. Note that both $g_i$'s and $h_i$'s depend on
the bosonic coordinates.  In the FL, as described above, it is possible
to decouple the spin degrees of freedom from the coordinate degrees of
freedom. For such cases, solving the supersymmetric Hamiltonian $H$, one
would in fact also be able to solve the corresponding spin chain problem.

\subsection{Systems related to short-range Dyson models}

A quantum Hamiltonian with nearest-neighbour and next-nearest-neighbour
inverse-square interactions among the particles was introduced
and studied in \cite{nni}. The model has relevance in the context of
random banded matrix theory describing short-range Dyson model\cite{rbm}.
In particular, the norm of the ground-state of this many-body system can be
identified with the joint-probability distribution function of the Gaussian
random banded matrix theory. Consequently, different correlation functions
of this Hamiltonian can be calculated exactly from the known results of
random banded matrix theory. The model has also relevance in the study
of nearest-neighbour spin chain models\cite{degu,spain1,spain3,spain4}.
A supersymmetric version of this nearest-neighbour analog of
rational $A_{N+1}$ Calogero Hamiltonian is constructed in this section.
The readers are referred to Ref. \cite{degu} for discussions on
a nearest-neighbour analog of rational $BC_{N+1}$ Calogero Hamiltonian.

The superpotential is chosen as,
\be
W= -\lambda \sum_{i=1}^N ln (x_i - x_{i+1}) + \frac{\omega}{2} \sum_{i=1}^N
x_i^2, \ \ \ \ x_{N+i}= x_i.
\label{eq16}
\ee
\noindent leading to the following expressions for $g_i$ and $h_i$,
\be
h_i = - \lambda (x_i - x_{i+1})^{-2}, \ \
g_i = \omega  - (h_i + h_{i-1}).
\label{eq16.1}
\ee
\noindent The Hamiltonian $H$ for the superpotential now reads,
\bea
H & = & - \frac{1}{2} \sum_{i=1}^N \frac{\partial^2}{\partial x_i^2} +
\frac{\lambda^2}{2} \sum_{i=1}^N \left [ 2 ( x_i- x_{i+1})^{-2}
- (x_{i-1}-x_i)^{-1} (x_i - x_{i+1})^{-1} \right ]\nonumber \\
& + & \lambda \sum_{i=1}^N \left [ ( x_i - x_{i+1} )^{-2} \left (
n_i + n_{i+1} -1
-\psi_i^{\dagger} \psi_{i+1} + \psi_i \psi_{i+1}^{\dagger} \right )
\right ]\nonumber \\
& + & \frac{1}{2} \omega^2 \sum_{i=1}^N x_i^2 + \omega \sum_{i=1}^N n_i -
\frac{N \omega}{2} - \lambda \omega N,
\label{eq17}
\eea
\noindent with the periodic boundary conditions on the fermionic variables:
$\psi_{N+i}= \psi_i$. Both nearest-neighbour and next-nearest-neighbour
interaction terms are present in the bosonic many-body potential. However,
only nearest-neighbour interaction terms are present for the fermions. This
allows a mapping of the bilinear terms involving fermionic operators in
terms of spin-spin interaction terms. The third term with the coefficient
$\lambda$ contains the XY Hamiltonian in terms of fermionic operators.
Thus, particles in this model are also having internal spin degrees of freedom.

The Hamiltonian $H$ reduces to the supersymmetric rational Calogero model 
for $N=3$ due to the periodic boundary conditions imposed on the bosonic and
the fermionic coordinates. The system is not exactly solved for a complete
set of states for $N >3$. However, infinitely many exact eigenstates can be
constructed analytically. An algebraic construction of these states in the
supersymmetric phase is described here. The ground state of $H$ in the
supersymmetric phase ( $\lambda > 0 $ ) is given by,
\be
\Phi_0 = \phi_0 |0 \rangle, \ \
\phi_0 \equiv \prod_i \left ( x_i - x_{i+1} \right )^{\lambda}
e^{-\frac{\omega}{2} \sum_i x_i^2}.
\label{eq17.0}
\ee
\noindent  The Hamiltonian $H$ can be mapped to a system of free
superoscillators satisfying the Eq. (\ref{map_osc}) with the
operator $S$ having the following expression:
\bea
S & = & \frac{1}{2} \sum_i \frac{\partial^2}{\partial x_i^2} +
\lambda \sum_i (x_i - x_{i+1})^{-1} \left ( \frac{\partial}{\partial x_i}
- \frac{\partial}{\partial x_{i+1}} \right )\nonumber \\
& - & \lambda \sum_i (x_i - x_{i+1})^{-2} \left [ \left ( \psi_i^{\dagger}
\psi_i+ \psi_{i+1}^{\dagger} \psi_{i+1} \right )
- \left ( \psi_i^{\dagger}
\psi_{i+1} - \psi_i \psi_{i+1}^{\dagger} \right ) \right ].
\label{eq17.02}
\eea
\noindent The construction of a partial set of eigen states from the free
superoscillator basis is given in Ref. \cite{degu}. Based on this analysis,
the following set of operators are introduced,
\bea
&& b_i^{-} =  i p_i = \frac{\partial}{\partial x_i}, \ \ \
b_i^{+}= 2 \omega x_i\nonumber \\
&& B_n^{-} = \sum_{i=1}^N T^{-1} b_i^{-^n} T, \ \ \
B_n^{+}=  \sum_{i=1}^N T^{-1} b_i^{{+}^n} T, \ \
n=1, 2, 3,\nonumber \\
&& F_m^{-}= \sum_i T^{-1} \psi_i b_i^{-^{m-1}} T,\ \ \
F_m^{+}=\sum_i T^{-1} \psi_i^{\dagger} b_i^{{+}^{m-1}} T, \ \
m=1, 2, 3, 4,\nonumber \\
&& q_n^{-}=T^{-1} \sum_i \psi_i^{\dagger} b_i^{-^n} T, \ \ \
q_n^{+}= T^{-1} \sum_i \psi_i b_i^{{+}^n} T, \ \
n=1, 2, 3,
\label{eq610}
\eea
\noindent which satisfy an algebra that reduces to (\ref{alg_osc})
for $\omega=1$. Unlike the case of rational Calogero model, there are
only three bosonic annihilation(creation) operators $B_n^{-}(B_n^{+})$ and
four fermionic annihilation(creation) operators $F_n^{-}(F_n^{+})$. 
The operators $B_n^{+}(F_m^+)$ lead to singular wave-function for
$n \geq 4 (m \geq 5)$ and are not acceptable.
The eigenstates can now be created in an algebraic manner by
using the above relations. In particular,
\be
\chi_{n_1 \dots n_{3} \nu_1 \dots \nu_4}= \prod_{k=1}^3 B_k^{+^{n_k}}
F_k^{+^{\nu_k}} F_4^{+^{\nu_4}} \ \Phi_0,
\label{eq6111}
\ee
\noindent is the eigenfunction with the eigen-value,
\be
E= \omega \left (\sum_{k=1}^{3} k (n_k + \nu_k) + 4 \nu_4 \right ).
\ee
\noindent  The bosonic quantum numbers $n_k$'s are
non-negative integers, while the fermionic quantum numbers $\nu_k$'s are
either $0$ or $1$. It appears that the eigen spectrum is not complete.

The Hamiltonian contains two independent parameters $\lambda$ and $\omega$.
The `freezing limit' is obtained by first scaling $\omega$ as $\lambda \omega$
in $H$ and then taking the limit $\lambda \rightarrow \infty$ of the operator
$\lambda^2 H$.  The spin degrees of freedom decouple completely from the
kinematic ones in this strong interaction limit with all the
particles frozen at their classical equilibrium configurations,
\be
W_i = \lambda \omega x_i + \lambda \left [ (x_{i-1} - x_i )^{-1} -
(x_i - x_{i+1})^{-1} \right ] = 0.
\label{eq17.1}
\ee
\noindent The XY Hamiltonian on a non-uniform lattice appears as a leading
term in this limit,
\be
H_{XY} = \sum_i \left [ ( \bar{x}_i - \bar{x}_{i+1} )^{-2} \left (
\frac{1}{2} ( n_i + n_{i+1} )
-\psi_i^{\dagger} \psi_{i+1} + \psi_i \psi_{i+1}^{\dagger} \right )
+ \frac{\omega}{2} n_i \right ],
\label{eq18}
\ee
\noindent where $\bar{x}_i$'s are determined from (\ref{eq17.1}). The general
solution of (\ref{eq17.1}) is not known. It is argued in Ref. \cite{degu}
that the equilibrium configurations of the particles necessarily constitute
a non-uniform lattice.

\section{Omitted topics, Open Arena and Summary}
\subsection{Omitted Topics}

(i) {\bf $SU(1,1|2)$ Supersymmetric Models in One Dimension}:\\

The motion of a test super-particle in the near-horizon limit of $3+1$
dimensional extremal Reissnar-Nordstr${\ddot{o}}$m black hole is suggested
to be described by the rational Calogero model with ${\cal{N}}=4$ $SU(1,1|2)$
superconformal symmetry\cite{gib}. A $2+1$ dimensional many-particle
system with extended $SU(1,1|2)$ superconformal symmetry is given in
Ref. \cite{pkg} and described in Section 6 of this article. However, such a
construction in space dimensions $D \neq 2$ meets with
difficulty\cite{nw,sc,belu,olaf2,olaf3,gala1,gala3,gala4,gala5}. The
rational $A_{N+1}$
Calogero model with $SU(1,1|2)$ supercnoformal symmetry is presented
in Ref. \cite{nw} for specific values of the strength of the inverse-square
interaction. The construction is based on $N$ bosonic and $4N$ fermionic degrees
of freedom and within this approach, the $SU(1,1|2)$ conformal symmetry is
absent for the generic values of the strength of the inverse-square
interaction\cite{nw}.  Several other attempts have been made to construct
${\cal{N}}=4$ superconformal models\cite{sc,belu,olaf2,olaf3,gala1,gala3,kiri}.
An important result in this context is described in Ref. \cite{kiri}
and an account of the whole development in this regard is given in
the review article \cite{review_scm}. However, an explicit construction
of the Hamiltonian for arbitrary number of particles and for the generic
values of the coupling constant(s), which contains the standard Calogero model
in its purely bosonic sector, is still beyond the reach.\\

\noindent (ii) {\bf Matrix Models, Collective Field theory etc.} :\\

An appropriate reduction of hermitian matrix model leads to rational
$A_{N+1}$ Calogero model\cite{poly}.
Similarly, a supersymmetric version of the Marinari-Parisi model can be
reduced to the supersymmetric rational $A_{N+1}$ Calogero model\cite{atish}.
Matrix model description of Calogero model appears in many other context,
for example, in Refs. \cite{gates,verlinde,sameer,agarwal}. The matrix models
related to Calogero Hamiltonian have also been studied in the large $N$ limit
through continuum collective theory techniques\cite{tonder,rodrigues,jevicki}.
Developments in this regard are described in the review
article \cite{jevicki}.\\

\noindent (iii) {\bf Superpolynomial}:\\

The zero fermion sector of the operator $H_1$ in Eq. (\ref{lasalle}) gives a
realization of the generalized Hermite polynomials. The orthogonal
eigenfunctions of the operator $H_1$ for any $N_f$ is known as generalized
hermitian superpolynomial\cite{sp2}. An algebraic construction of
superpolynomials related to supersymmetric Calogero-Sutherland systems are
described in Refs. \cite{sp0,sp1,sp2,sp3,sp4}. The integrable structure
of these supersymmetric systems are also explored. 

\subsection{Open Arena}

(i) {\bf Self-adjoint Extensions}:\\

The self-adjoint extension of rational
Calogero model has been studied in Refs. \cite{bbm,tsutsui}. The scale
invariance of the rational Calogero model with purely inverse-square
interaction gets broken at the quantum level due to the imposition of modified
boundary conditions and the Hamiltonian admits bound states. Similarly, this
new quantization scheme for the rational Calogero model with the harmonic
confinement term leads to non-equispaced energy levels with a negative energy
bound state\cite{bbm}. The supersymmetric Hamiltonian (\ref{eq6.2}) reduces
to the Calogero model in the zero fermion sector. The results stated
above are thus equally valid in the zero fermion sector. A systematic study
on self-adjoint extensions of the supersymmetric rational Calogero model
for arbitrary $N_f$ is desirable.\\

\noindent (ii) {\bf `Atypical' Calogero Model}:\\

The construction of the rational Calogero Hamiltonian with $OSp(2|2)$
supersymmetry is not unique. The standard Hamiltonian (\ref{eq6.2}) or
(\ref{hami_bn}) corresponds to the `typical' representation. A new
Hamiltonian corresponding to the `atypical' representation has been
introduced in Eq. (\ref{h2}). The complete spectrum and the integrable
structure of this Hamiltonian is not known. A set of eigen values and the
eigenfunctions of this Hamiltonian corresponding to the underlying $OSP(2|2)$
symmetry may be obtained analytically. The quadratic and the cubic Casimir
operators being zero in the `atypical' representation, these operators can not
be used to characterize the spectra. Further, as in the case of `typical'
Calogero model, the spectrum generating algebra is expected to be larger
than $OSp(2|2)$. This expectation stems from the fact that superoscillator
Hamiltonian in the `atypical' representation of $OSp(2|2)$ contains a
spin-orbit interaction term and its spectrum is different from the
superoscillator model corresponding to `typical' representation. It may be
noted that a non-trivial mixing of angular momentum operators $L_{ij}$ and
$J_{ij}$ also appears in the `atypical' Calogero Hamiltonian (\ref{eq6.2}).
A study on the exact solvability of the `atypical' model may reveal some
of its hidden surprises.\\

\noindent (iii) {\bf Generalized Calogero-type Models}:\\

There are many supersymmetric systems with inverse-square interactions for
which only a part of the complete spectrum can be obtained analytically. Two
such physically relevant systems are presented in this review article
in sections 6 and 7.  Any supersymmetric Hamiltonian corresponding to
the `typical' representation of $OSp(2|2)$ can be mapped to a system of
free superoscillators through a similarity transformation. However, in general,
the spectrum of the original many-body Hamiltonian is not identical with
that of superoscillators. It appears that the similarity operator takes
the many-body Hamiltonian out of its Hilbert space. The choice of the
free superoscillator basis respecting the discrete symmetries of the 
rational $A_{N+1}(B_{N+1})$ Calogero model gives the complete
eigen spectrum of the model. Further investigations on hidden symmetries,
if any,
of the generalized Calogero-type models are desirable. In general, the nature
of the similarity operator and its action on free superoscillator basis needs
further investigation for a better understanding of generalized Calogero-type
models.\\

\noindent (iv) {\bf Pseudo-hermitian supersymmetric Calogero Models}:\\

The deformation of the rational $A_{N+1}$ Calogero model without the confining
term lead to a pseudo-hermitian supersymmetric system with  broken
translational invariance. The configuration space of the original Hamiltonian
is different from the deformed Hamiltonian and the eigen spectra of these
models become identical, only when solved for identical boundary conditions.
The deformed model with allowed modified boundary condition(s) is expected to
have different spectra and need further investigations.

A Coxter-invariant superpotential was constructed in Ref. \cite{sasa} leading
to a unified description of supersymmetric Calogero-Moser-Sutherland models
based on all the root systems with the rational, trigonometric and hyperbolic
potentials. Such a universal description of pseudo-hermitian supersymmetric
Calogero models presented in this review article is desirable. 
Further, pseudo-hermitian non-supersymmetric Calogero models have
been constructed by considering complex root spaces that are invariant under
anti-linear involutions related to all Coxter groups\cite{smith}. The
deformations considered in Ref. \cite{smith} involve discrete transformations
in the root space, while only continuous deformation was considered in section
5. A construction of supersymmetric version of the models considered in Ref.
\cite{smith} is desirable.

\subsection{Summary}

The main results presented in this article are based on previously
published works\cite{susy,nonu,me,pkg,piju_latest,degu} and may be summarized
as follows.

\begin{itemize}

\item The condition of shape invariance can be used in conjunction with
Dunkl operator to obtain the complete eigen spectrum of the rational
Calogero model, i.e. the Hamiltonian appearing in the $N_f=0$ sector
of the supersymmetric Hamiltonian.

\item The rational Calogero model in the supersymmetry-preserving
phase can be mapped to a set of free superoscillators through a similarity
transformation. The complete eigen spectrum of the Calogero model can be
constructed from those eigenstates of the free superoscillator Hamiltonian
which are invariant under the discrete symmetries of the many-body parent
Hamiltonian.

\item The supersymmetry-breaking phase of the rational Calogero model
can also be studied by mapping a dual Hamiltonian to free superoscillator
Hamiltonian. All the eigenstates in the supersymmetry-breaking phase can
be constructed from permutation-symmetric superoscillator basis via
the dual Hamiltonian.

\item A `necessary condition' for the equivalence of a many-body
supersymmetric Hamiltonian with $OSp(2|2)$ symmetry to a system of free
super-oscillators is that the Hamiltonian should commute with the total
fermion number operator $N_f$. The proof of `sufficient condition' is model
dependent and nontrivial. The equivalence can be proved for rational Calogero
models.

\item The super-extension of rational Calogero model with $OSp(2|2)$
supersymmetry is not unique. A new Hamiltonian corresponding to the
`atypical' representation of $OSp(2|2)$ has been constructed. The quadratic and
the cubic Casimir operators vanish identically and can not be used to
characterize the spectrum. It is not known whether the `atypical' Calogero
model is integrable or not.

\item A construction of pseudo-hermitian supersymmetric Calogero model has
been given, which is isospectral with the standard Calogero model. This
construction is valid for rational, trigonometric or hyperbolic versions
of Calogero models and also for any root system.

\item A $2+1$ dimensional many-body system with extended ${\cal{N}}=4$
$SU(1,1|2)$ superconformal symmetry has been presented for the generic values
of the coupling constant as well as for arbitrary number of particles.

\item A supersymmetric Hamiltonian describing $N$ spinless particles
with nearest-neighbour and next-nearest-neighbour inverse-square
interaction has been shown to be equivalent to a system of interacting
bosonic particles with internal spin degrees of freedom. Models of XX
spin chains may be obtained from this supersymmetric Hamiltonian in
appropriate limits.
\end{itemize}

The findings of this study are relevant in enriching a general understanding
of the integrable structure of many-particle supersymmetric quantum systems.
Further studies may reveal new avenues on the applicability of the
mathematical techniques related to shape-invariance and supersymmetry
to generic quantum systems with more than one bosonic and one fermionic
degrees of freedom. Further, the conformal symmetry is a recurrent theme
in many areas like black-holes, matrix models, string theory, strongly
correlated system etc. and offers universal description of many apparently
diverse physical systems. The superconformal systems studied in this review
thus have potential applications in diverse subjects, including a
possible futuristic realization of many-body quantum systems with both
spatial and internal degrees of freedom in the laboratory.

\section{Acknowledgment}

The Author would like to thank B. Basu-Mallick, G. Date, T. Deguchi, Kumar
S. Gupta, A. Khare, S. P. Khastgir, M. V. N. Murthy, R. Sasaki and M. Sivakumar
for discussions on the topic at various points of time and contributing to his
understanding of the subject.

\section{Appendices}

A few mathematical results which are relevant in the discussions of the
main text are discussed in sections 9.1, 9.2, 9.3 and 9.5. The rational
$BC_{N+1}$ Calogero model has not been included in the main text. The
relevant discussions in this regard are included in Appendix-IV in
section 9.4.

\subsection{Appendix-I: Matrix representation of the real Clifford algebra}

A $2^N \times 2^N$ matrix representation of the elements of the Clifford
algebra (\ref{cliff}) may be given in terms of the Pauli matrices
$\sigma^a, a=1,2,3$ and the $2 \times 2$ identity matrix $I$ as follows,
\bea
&& \xi_1 = \sigma^1 \otimes I \otimes I \otimes \dots \otimes I, \hspace{.8in} 
\xi_{N+1} = \sigma^2 \otimes I \otimes I \otimes \dots \otimes I,\nonumber \\
&& \xi_2 = \sigma^3 \otimes \sigma^1 \otimes I \otimes \dots \otimes I,
\hspace{.7in}
\xi_{N+2} = \sigma^3 \otimes \sigma^2 \otimes I \otimes \dots \otimes
I,\nonumber \\
&& \vdots \hspace{2.35in} \vdots \nonumber \\
&& \xi_i = \sigma^3 \otimes \dots \otimes \sigma^3 \otimes \sigma^1
\otimes I \otimes \dots \otimes I,\hspace{.1in}
\xi_{N+i} = \sigma^3 \otimes \dots \otimes \sigma^3 \otimes \sigma^2
\otimes I \otimes \dots \otimes I,\nonumber \\
&& \vdots \hspace{2.35in} \vdots \nonumber \\
&& \xi_{N-1} = \sigma^3 \otimes \dots \otimes \sigma^3 \otimes \sigma^1
\otimes I,\hspace{.5in}
\xi_{2N-1} = \sigma^3 \otimes \dots \otimes \sigma^3 \otimes \sigma^2
\otimes I,\nonumber \\
&& \xi_N = \sigma^3 \otimes \sigma^3 \otimes \dots \otimes \sigma^3
\otimes \sigma^1,\hspace{.55in}
\xi_{2N} = \sigma^3 \otimes \sigma^3 \otimes \dots \otimes \sigma^3
\otimes \sigma^2.
\eea
\noindent The matrices $\xi_p$ are hermitian. The matrices $\xi_i$ are
symmetric, while $\xi_{N+i}$ are anti-symmetric for any $i$. The fermionic
operators $\psi_i$ and $\psi_i^{\dagger}$ have the matrix representation,
\bea
\psi_i & = & \sigma^3 \otimes \dots \otimes \sigma^3 \otimes \sigma_- \otimes I
\otimes \dots \otimes I\nonumber \\
\psi_i^{\dagger} & = & \sigma^3 \otimes \otimes \dots \sigma^3 \otimes \sigma_+
\otimes I \otimes \dots \otimes I,
\eea
\noindent where $\sigma_{\pm}=\frac{1}{2}
\left ( \sigma^1 \pm i \sigma^2 \right )$ are in the $i^{th}$ position,
preceded by the tensor-product of $i-1$ numbers of $\sigma^3$ and followed
by the tensor-product of $N-i$ numbers of the matrix $I$.

\subsection{Appendix-II:$OSp(2|2)$ superalgebra}

The structure-equations of the $OSp(2|2)$ superalgebra are
described in terms of a set of fermionic generators
$f \equiv \{Q_1, Q_2, S_1, S_2 \}$ and a set of bosonic generators
$b \equiv \{ H, D, K, Y \}$ as follows\cite{superlie,rev}:
\bea
&& \{Q_{\alpha}, Q_{\beta} \} = 2 \delta_{\alpha \beta} H, \ \
\{S_{\alpha}, S_{\beta}\} = 2 \delta_{\alpha \beta} K, \ \
\{Q_{\alpha},S_{\beta}\} = - 2 \delta_{\alpha \beta} D + 2 \epsilon_{\alpha
\beta} Y,\nonumber \\
&& [H, Q_{\alpha}]=0, \ [H,S_{\alpha}]=-i Q_{\alpha}, \
[K,Q_{\alpha}] = i S_{\alpha}, \ [K,S_{\alpha}]=0,\nonumber \\
&& [D, Q_{\alpha}] = - \frac{i}{2} Q_{\alpha}, \
[D, S_{\alpha}] = \frac{i}{2} S_{\alpha}, \
[Y,Q_{\alpha}] = \frac{i}{2} \epsilon_{\alpha \beta}
Q_{\beta}, \ \ [Y, S_{\alpha}]=\frac{i}{2} \epsilon_{\alpha \beta}
S_{\beta},\nonumber \\
&& [Y,H]=[Y,D]=[Y,K]=0, \ \ \alpha, \beta = 1, 2.
\label{ap_II.1}
\eea
\noindent The bosonic operators $H$, $D$ and $K$ generate the $O(2,1)$
sub-algebra of $OSp(2|2)$,
\be
[H,D] = i H, \ \ [H,K] = 2 i D, \ \ [D,K] = i K,
\label{ap_II.2}
\ee
\noindent with its Casimir operator having the form,
\be
C=\frac{1}{2} ( H K + K H ) - D^2.
\label{ap_II.3}
\ee
\noindent The quadratic and the cubic Casimir operators of
$OSp(2|2)$ are given by\cite{superlie,rev},
\bea
&& C_2= C + \frac{i}{4} [Q_1,S_1] + \frac{i}{4} [Q_2,S_2] - Y^2,\nonumber \\
&& C_3= C_2 Y - \frac{Y}{2} + \frac{i}{8} {\Big (} [Q_1,S_1] Y +
[Q_2,S_2] Y + [S_1,Q_2] D\nonumber \\
&& \ \ \ \ - [S_2,Q_1] D + [Q_1,Q_2] K + [S_1,S_2] H  {\Big )}.
\label{casi1}
\eea
\noindent Both `typical' and `atypical' representations of the supergroup
$OSp(2|2)$ are allowed. The quadratic and the cubic Casimir operators vanish
identically for the `atypical' representation of the group.

The subgroup $OSp(1|1)$ of $OSp(2|2)$ is described either by the
set of generators $A_1 \equiv \{ H, D, K, Q_1, S_1 \}$ or
$A_2 \equiv \{ H, D, K, Q_2, S_2 \}$. The Casimir of the $OSp(1|1)$
corresponding to the set $A_1$ is given by,
\be
C_1 = C + \frac{i}{4} [Q_1, S_1] + \frac{1}{16}.
\ee
\noindent An even operator $C_s$, known as Scasimir, may be defined as follows,
\be
C_s = i [Q_1, S_1] - \frac{1}{2},
\label{s-casi}
\ee
\noindent which has the property that it commutes with all the
bosonic generators and anti-commutes with all the fermionic generators
of the set $A_1$. The Scasimir $C_s$ is related to the Casimir operators
$C$ and $C_1$ through the relations,
\be
C_1= \frac{1}{4} C_s^2, \ \ 
C=\frac{1}{4} C_s ( C_s-1) - \frac{3}{16}.
\ee
\noindent The Casimir and the Scasimir for the set $A_2$ are given by,
\be
\bar{C}_1 = C + \frac{i}{4} [Q_2, S_2] + \frac{1}{16}, \ \
\bar{C}_s = i [Q_2, S_2] - \frac{1}{2},
\label{bar_s-casi}
\ee
\noindent which satisfy the identities
\be
\bar{C}_1=\frac{1}{4} \bar{C}_s^2, \ \
C=\frac{1}{4} \bar{C_s} (\bar{C_s}-1) - \frac{3}{16}.
\label{ccseq}
\ee
\noindent This implies that, in general, the Casimir $C$ can be factorized in
two different ways, either in terms of $C_s$ or $\bar{C}_s$.

\subsection{Appendix-III: $SU(1,1|2)$ superalgebra}

The structure equations of $SU(1,1|2)$ superalgebra are described in terms
of a set of fermionic generators (${\cal{Q}}_p, {\cal{S}}_p,
{\cal{Q}}_p^{\dagger}, {\cal{S}}_p^{\dagger}$),  a set of bosonic
generators ($ J_a, H, D, K$) and a central element $T$ as follows:
\bea
&& \left \{ {\cal{Q}}_p, {{\cal{Q}}^{\dagger}}^r \right \} = 2 \delta_p^r H,
 \ \ \left \{ {\cal{S}}_p, {{\cal{S}}^{\dagger}}^r \right \} =
2 \delta_p^r K, \ \
\left \{ {\cal{Q}}_p, {{\cal{Q}}}_r \right \} = 0, \ \
\left \{ {\cal{S}}_p, {{\cal{S}}}_r \right \} = 0, \nonumber \\
&& \left \{ {\cal{Q}}_p, {{\cal{S}}^{\dagger}}^r \right \} = 2 i
\left ( \sigma^a \right )_p^r J_a - 2 \delta_p^r D - i \delta_p^r T, \ \
\left [ J_a, {\cal{Q}}_p \right ] = -\frac{1}{2} \left ( \sigma_a \right )_p^r
{\cal{Q}}_r,\nonumber \\
&& \left \{ {\cal{S}}_p, {{\cal{Q}}^{\dagger}}^r \right \} = - 2 i
\left ( \sigma^a \right )_p^r J_a - 2 \delta_p^r D + i \delta_p^r T, \ \
\left [ J_a, {\cal{S}}_p \right ] = -\frac{1}{2} \left ( \sigma_a \right )_p^r 
{\cal{S}}_r,\nonumber \\
&& \left [ {\cal{Q}}_p, D \right ] = \frac{i}{2} {\cal{Q}}_p, \ \
\left [ {{\cal{Q}}^{\dagger}}^p, D \right ] =
\frac{i}{2} {{\cal{Q}}^{\dagger}}^p,
\left [ {\cal{S}}_p, D \right ] = - \frac{i}{2} {\cal{S}}_p,
\left [ {{\cal{S}}^{\dagger}}^p, D \right ] =
-\frac{i}{2} {{\cal{S}}_p^{\dagger}},\nonumber \\
&& \left [ K, {\cal{Q}}_p \right ] = i {\cal{S}}_p, \ \
\left [ K, {{\cal{Q}}^{\dagger}}^p \right ] = i {{\cal{S}}^{\dagger}}^p,
\ \ \left [ H, {\cal{S}}_p \right ] = - i {\cal{Q}}_p, \ \
\left [ H, {{\cal{S}}^{\dagger}}^p \right ] = - i {{\cal{Q}}^{\dagger}}^p,
\nonumber \\
&& \left [J_a, J_b \right ] = i \epsilon_{abc} J_c, \ \
p, r=1, 2, \ \ a, b, c=1, 2, 3.
\eea
\noindent The bosonic operators $H$, $D$ and $K$ generate the $O(2,1)$
algebra given in Eq. (\ref{ap_II.2}). A summation over the repeated indices
is implied in the above expressions.

\subsection{Appendix-IV: Rational $BC_{N+1}$ Calogero Model}

The superpotential for the rational $BC_{N+1}$-type Calogero model
has the following expression,
\bea
W (\lambda, \lambda_1, \lambda_2) & = & - \lambda \sum_{i <j=1}^N
ln \ \left ( x_i^2 - x_j^2 \right ) - \sum_{i=1}^N \left [
\lambda_1 ln \ x_i + \lambda_2 ln (2 x_i) \right ]\nonumber \\
& + & \frac{\omega}{2} \sum_{i=1}^N x_i^2,
\label{sp_bn}
\eea
\noindent where $\lambda$, $\lambda_1$ and $\lambda_2$ are arbitrary
parameters. The $D_{N+1}$-type model is described by
$\lambda_1=\lambda_2=0$, while $\lambda_1=0 (\lambda_2=0)$ describes
$C_{N+1} (B_{N+1})$-type Hamiltonian. The discussion in this article is
restricted to the $B_{N+1}$-type Calogero model for which the Hamiltonian
is given by,
\bea
H_{B_{N+1}} & = & - \frac{1}{2} \sum_i \frac{\partial^2}{\partial x_i^2}
+ \frac{\omega^2}{2} \sum_i x_i^2 +
\frac{1}{2} \lambda (\lambda-1) \sum_{i \neq j} \left [ x_{ij}^{-2}
+ (x_i + x_j )^{-2} \right ]\nonumber \\
& + & \frac{1}{2} \lambda_1 ( \lambda_1-1) \sum_i x_i^{-2}
+ \omega \sum_i \psi_i^{\dagger} \psi_i  + \lambda_1 \sum_i \psi_i^{\dagger}
\psi_i x_i^{-2}\nonumber \\
& + & \lambda \sum_{i \neq j} \left [ x_{ij}^{-2}
\left ( \psi_i^{\dagger} \psi_i - \psi_i^{\dagger} {\psi_j} \right ) +
(x_i + x_j )^{-2} \left ( \psi_i^{\dagger} \psi_i +
\psi_i^{\dagger} {\psi_j} \right ) \right ]\nonumber \\
& - & \frac{\omega}{2}  N \left [  1 + 2 \lambda (N-1) + \lambda_1 \right ].
\label{hami_bn}
\eea
\noindent It may be noted that the many-body inverse-square interaction is
not transnational invariant. Apart from the transnational invariant
mutual inverse-square interaction between any pair of particles, each particle
also interacts with the images of all other particles and also with itself.
This kind of Hamiltonians are suitable for describing systems with boundaries.
The many-body wave-functions are taken to be vanishing at the singular points
$x_i=0, x_i = \pm x_j \forall i,j$ and the eigen-value problem is solved in
the $0 < x_1 < x_2 < \dots < x_N$ sector of the configuration space. The
Hamiltonian is invariant under the permutation symmetry ${\cal{P}}$ and the
reflection symmetry ${\cal{R}}$:
\bea
&& {\cal{P}}: \ \ x_i \rightarrow x_j, \ \ \psi_i \rightarrow \psi_j,
\ \ \psi_i^{\dagger} \rightarrow \psi_j^{\dagger}\nonumber \\
&& {\cal{R}}: \ \ x_i \rightarrow - x_i, \ \ \psi_i \rightarrow - \psi_i,
\ \ \psi_i^{\dagger} \rightarrow - \psi_i.
\label{dis_bn}
\eea
\noindent These two symmetries allow a smooth continuation of the
wave-functions from a given sector of the configuration space to all other
sectors. The reflection symmetry also has an interesting consequence on the
spectrum. The ground-state of the Hamiltonian in the supersymmetric phase is
given by,
\be
\Phi_0= \phi_0 |0\rangle, \ \
\phi_0 \equiv \prod_{i <j} \left ( x_i^2 - x_j^2 \right )^{\lambda}
\prod_k x_k^{\lambda_1}  e^{- \frac{1}{2} \sum_i x_i^2},
\ \ \lambda, \lambda_1 > 0,
\label{gr_bn}
\ee
\noindent which is normalizable for $\lambda, \lambda_1 > -\frac{1}{2}$.
However, the positivity of $\lambda$ and $\lambda_1$ is imposed so that
each momentum operator $p_i$ is self-adjoint for the wave-function of the
form $\Phi_0$.

\subsubsection{Shape Invariance \& Exact Solvability}

The $N_f=0$ sector of the the supersymmetric Hamiltonian
is described by the bosonic Hamiltonians ${\cal{H}}^{(0)}$:
\bea\label{hami_bcn_0}
{\cal{H}}^{(0)}(\lambda, \lambda_1, \omega) & = & 
{\cal{H}}^{B_{N+1}} - E_0^{B_{N+1}}\nonumber \\
{\cal{H}}^{B_{N+1}} & := & {\frac{1}{2}}{\sum_{i} {p_{i}^2}}
+ \frac{1}{2} \lambda(\lambda - 1) \sum_{i \neq j} \left [ \left ( x_{i} -
x_{j} \right ) ^{-2} + { \left ( x_{i} +x_{j} \right )} ^{-2} \right
]\nonumber \\
& + & \frac{1}{2} \lambda_{1} (\lambda_{1}-1) \sum_{i=1}^N x_i^{-2}
+ {\frac{\omega}{2}}  {\sum_{i} x_i^2}\nonumber \\  
E_0^{B_{N+1}} & \equiv & \frac{\omega}{2}  N \left [  1 + 2 \lambda (N-1) +
\lambda_1 \right ].
\eea
\noindent The bosonic Hamiltonian ${\cal{H}}^{(N)}$ corresponding to the
$N_f=N$ sector of the supersymmetric Hamiltonian is related to
${\cal{H}}^{(0)}$ through the shape-invariance condition,
\be
{\cal{H}}^{(N)}(\lambda, \lambda_1, \omega) = {\cal{H}}^{(0)}(\lambda + 1,
\lambda_1 + 1, \omega ) + \frac{N \omega}{2} \left ( 2 N + 1 \right ).
\ee
\noindent However, the complete spectra of ${\cal{H}}^{(0)}$ and
${\cal{H}}^{(N)}$ can not be obtained by using this shape-invariance condition.
This is because of the absence of intertwining relations between these two
Hamiltonians which appear in the description of supersymmetric system with
one bosonic and one fermion degrees of freedom. The present situation is
circumvented by the use of a modified shape-invariance condition and exchange
operator formalism as applied to the system with $B_{N+1}$-type Hamiltonian.

The Dunkl operator is introduced in terms of the exchange operator $M_{ij}$
and the reflection operator $t_i$:
\be\label{2.2}
{\cal{D}}_{i}
= -i \partial_{i} + i \lambda \sum_{j(\neq i)} \left [ {(x_i -x_j)}^{-1} M_{ij}
+ {(x_i +x_j)}^{-1} \tilde {M_{ij}} \right ] +i \lambda_1 x_{i} ^{-1},\ \ \
\tilde{M_{ij}} := t_{i}t_{j}M_{ij}~.
\ee
\noindent
The reflection operator $t_i$ satisfies the following relations,
\bea
&& t_i x_j = x_j t_i \ for \ i \neq j, \ \ t_i  x_i +x_i t_i =0 \
\forall \ i, \ \ t_i^2=1 \ \forall \ i,\nonumber \\
&& M_{ij} t_i = t_j M_{ij}~,\ \
\tilde{M}_{ij}^\dagger = \tilde{M}_{ij}~,\ \
t_i {\cal{D}}_i = - {\cal{D}}_{i} t_i~,\ \
t_i {\cal{D}}_j = {\cal{D}}_j t_i~ \ for j \neq i,\nonumber \\
&& \tilde {M_{ij}} {\cal{D}}_i = -{\cal{D}}_{i} \tilde{M_{ij}}, 
\ t_i \phi(x_1, \dots, x_i, \dots, x_N)=
\phi(x_1, \dots, -x_i, \dots, x_N).
\eea
\noindent The Dunkl operators ${\cal{D}}_i$ commute among themselves
and the following identity holds:
\bea
&& b_i  := {\cal{D}}_i - i \omega x_i, \ \ b_i^{\dagger}:= {\cal{D}}_i
+ i \omega x_i,\nonumber \\
&& [{{b}}_i, {{b}}_j^\dagger] = 2 \omega \delta_{ij} \left (1+
\lambda \sum_{k (\neq i)} ( M_{ik}+
\tilde{M_{ik}}) +2 \lambda_{1}t_{i} \right )\nonumber \\
&& \ \ \ \ \ \ \ \ \ \ \ \ - 2(1- \delta_{ij}) \lambda \omega (M_{ij} - \tilde{M_{ij}})~.
\eea
\noindent All other commutators involving $b_i$ and their adjoints vanish
identically. The supersymmetric partner Hamiltonians ${\cal{H}}$ and
$\tilde {{\cal{H}}}$ for the $B_{N+1}$ case may be defined in a similar
way to that of rational $A_{N+1}$ Calogero model,
\be
{\cal{H}} = \frac{1}{2}\sum_{i=1}^N {{b}}_i^\dagger {{b}}_i \, ~, \ \
\tilde{{\cal{H}}} = \frac{1}{2}\sum_{i}{{b}}_i {{b}}_i^\dagger.
\ee
\noindent The Hamiltonian ${\cal{H}}$ reduces to $H^{(0)}$ in Eq.
(\ref{hami_bcn_0}) if $M_{ij}$ acts on symmetric functions,
whereas it reduces to $H^{(N)}$ $M_{ij}$ is restricted to the
subspace of anti-symmetric functions. A set of annihilation and
creation operators, similar to $A_n, A_n^{\dagger}$ in Eq. (\ref{ani1}),
are introduced as follows:
\be
\hat{B}_{n} = \sum_{i=1}^N {{b}}_{i}^2 \, ~, \ \
\hat{B}_{n}^\dagger = \sum_{i=1}^N ({b}_i^\dagger)^2 \, \ n \leq N.
\ee
\noindent It can be shown that if ${{\phi}} (\tilde {{{\phi}}})$ is the
eigenfunction of
${\cal{H}} (\tilde{{\cal {H}}})$ with eigenvalue ${\cal{E}}
(\tilde {{\cal{E}}})$ then
\be
{\cal{H}}(\hat{B_{2}}^\dagger {\phi}) =
(\tilde{{\cal{E}}}-\hat{\delta}_{2})(\hat{B_{2}}^\dagger
\tilde {\phi})~,\ \
\tilde {{\cal{H}}}(\hat{B}_{2}\phi) =
({\cal{E}}+\hat{\delta}_{2})(\hat{B}_{2}\phi)~~.
\ee
\noindent where,
\be
\hat{\delta}_{2} = [N-2 \pm 2 \lambda N(N-1) +2 \lambda_{1}N ] \omega~.
\ee
\noindent It may be noted that he operator which brings in a correspondence
between the eigenstates $\phi$ and $\tilde{\phi}$ is $B_2$, not $B_1$.
This is because the reflection symmetry of the $BC_{N+1}$ Hamiltonian 
is also a symmetry of the wave-functions provided the operator
${B}_2$ is used instead of ${B}_1$. The shape invariance condition for
$H$ and $\tilde {H}$ of the $B_N$ model reads,
\bea
&& \tilde {H} (\lambda,\lambda_1,\omega) = H (\lambda,
\lambda_1,\omega)
+ R_2 (\lambda,\lambda_1,\omega)\nonumber \\
&& R_{2}(\lambda,\lambda_1,\omega) \equiv \left [ N  \pm 2 \lambda N \left ( N
- 1 \right ) + 2 \lambda_{1} N \right ] \omega~~.
\eea
\noindent Following the standard treatment, the spectrum of $H^{(0)}$
is determined as,
\be
E_{n}=n(R_{2}-\hat{\delta}_{2}) =2n\omega~.
\ee
\noindent  The reflection symmetry is manifested in the spectrum which now
depends on  $2 n \omega$, instead of $n\omega$ as in the case of
rational ${A}_{N+1}$ Calogero model. The eigen functions are obtained by acting
the operators $B_n$ on the ground state $\phi_0$ of the $B_{N+1}$ model, as in
Eq. (\ref{28}) for $A_{N+1}$ model.

\subsubsection{Mapping to free super-oscillators}

The supersymmetric rational $B_{N+1}$-type Calogero Hamiltonian in
Eq. (\ref{hami_bn}) can be mapped to a system of $N$ free superoscillators
by using Eq. (\ref{map_gen}) with $W(\lambda, \lambda_1, \lambda_2=0)$
given by Eq.(\ref{sp_bn}). The supersymmetric and the supersymmetry-breaking
phases are discussed separately.\\

\noindent {\bf Case I: Supersymmetric phase: $\lambda, \lambda_1 >0$}:\\

\noindent The construction of the eigen functions requires the following
choice of the free superoscillator basis:
\be
\hat{P}_{n,k} = \frac{1}{N_f!} r^{2 n}
\sum_{i_1, i_2, \dots, i_{N_f}} f_{i_{1} i_{2} \dots i_{N_f}}
(x_{i_1} \psi_{i_1}^{\dagger})
(x_{i_2} \psi_{i_2}^{\dagger}) \dots (x_{i_{N_f}} \psi_{i_{N_f}}^{\dagger}),
\label{eq15}
\ee
\noindent where $f$ is anti-symmetric under the exchange of any two indices
and a homogeneous function of degree $2 k$ of the bosonic coordinates.
The functions $\hat{P}_{n,k}$
are invariant under the discrete symmetry (\ref{dis_bn}). The action of
the similarity operator $\tilde{T}$ on functions without the discrete
symmetry (\ref{dis_bn}) produces essential singularity in the wave-functions
which are not physically acceptable. Thus, the complete spectrum of
$H_{B_{N+1}}$ is described by a subset of the spectrum of super-oscillators,
\be
E_{B_{N+1}}= 2 ( n + k + N_f ), \ \  E_{sho}= 2 n + k + N_f.
\label{eq14}
\ee
\noindent The rational $A_{N+1}$ Calogero model is equivalent to a set
of $N$ free super-oscillators, whereas the rational $B_{N+1}$ Calogero
model is equivalent to a set of $N$ free `super-half-oscillators'\cite{susy}.

The eigen-spectrum can be constructed in an algebraic way by
defining the creation and annihilation operators as,
\be
{\tilde{B}}_{n}^+ = {\tilde{T}}^{-1} \sum_i b_i^{+^{2 n}} {\tilde{T}}, \ \
{\tilde{F}}_{n}^+ = {\tilde{T}}^{-1} \sum_i \psi_i^{\dagger}
b_i^{+^{2 n-1}} {\tilde{T}}, \ \
\label{eqq}
\ee
\noindent which are invariant under the discrete symmetry (\ref{dis_bn}). Thus,
the eigen-functions obtained by operating these operators on the ground-state
$ \Phi_0$ in Eq. (\ref{gr_bn}) are also invariant under the same discrete
symmetry. The eigen-states are obtained as,
\be
{\cal{\chi}}_{n_1 \dots n_{N} \nu_1 \dots \nu_{N}}=
\prod_{k=1}^N {\tilde{B}}_k^{+^{n_k}}
{\tilde{F}}_k^{+^{\nu_k}} \ \Phi_0.
\label{eq66}
\ee
\noindent with the energy ${\cal{E}}= \sum_{k=1}^{N} 2 k (n_k + \nu_k)$.
The bosonic quantum numbers are non-negative integers, while the fermionic
quantum numbers are $0$ or $1$. \\

\noindent {\bf Case II: Supersymmetry-breaking phase}:\\

\noindent There are three regions in the parameter space for which the
supersymmetry is broken: (i) $\lambda < 0$, $\lambda_1 < 0$, (ii)
$\lambda < 0$, $\lambda_1 > 0$ and (iii) $\lambda > 0$, $\lambda_1 < 0$.
The eigen-spectrum of the Hamiltonian in the region (i) can be obtained in a
similar way as in the case of supersymmetry-breaking phase of the rational
$A_{N+1}$ Calogero model. However, modified treatments are required for
obtaining the spectra in the regions (ii) and (iii). The readers are referred
to Ref. \cite{susy} for details. The eigen spectra in these three regions are
given below:
\bea
&& \lambda < 0, \lambda_1 < 0: \ \ E= N \left [ 1 - 2 \lambda (N-1) -
\lambda_1 \right ] + E_{n_k,\nu_k}\\
&& \lambda <  0, \lambda_1 > 0: \ \ E= N \left [ \frac{3}{2} -
2 \lambda (N-1) \right ] + E_{n_k, \nu_k}\\
&& \lambda > 0, \lambda_1 < 0: \ \
E= N \left [ \frac{3}{2} - 2 \lambda_1 (N-1) \right ] + 
E_{n_k,\nu_k},
\eea
\noindent where $E_{n_k,\nu_k} = \sum_{k=1}^N 2 \left [ k n_k +
(k-1) \nu_k \right ]$.
The bosonic quantum numbers $n_k$'s are non-negative integers,
while the fermionic quantum numbers $\nu_k=0,1$.

\subsubsection{ Nearest-neighbour variant of rational $BC_{N+1}$ Calogero model}

The Hamiltonian $H$ in terms of the variables
$q_i = x_i - x_{i+1}, \ \ \bar{q}_i = x_i + x_{i+1}$ reads,
\bea
H & = & \frac{1}{2} \sum_i p_i^2 + \lambda^2 \sum_i \left [
q_i^{-2}+ \bar{q}_i^{-2} - (q_{i-1}^{-1} - \bar{q}_{i-1}^{-1})
(q_i^{-1} + \bar{q}_i^{-1}) \right ]\nonumber \\
& - & \lambda \omega \left (2 + \frac{\lambda_1}{\lambda} \right) N
+ \frac{\lambda}{2} \sum_i q_i^{-2}
\left [ (n_i + n_{i+1} )-
2 \left (\psi_i^{\dagger} \psi_{i+1} - \psi_i \psi_{i+1}^{\dagger} \right )
\right ]\nonumber \\
& + & \frac{\lambda}{2} \sum_i \bar{q}_i^{-2}
\left [ (n_i + n_{i+1} ) + 2 \left (\psi_i^{\dagger} \psi_{i+1} -
\psi_i \psi_{i+1}^{\dagger} \right ) \right ]\nonumber \\
& + &
\frac{1}{2} \sum_i \left [ \left ( \omega +
\frac{\lambda_1}{x_i^2} \right ) n_i 
+ \omega^2 x_i^2 + \frac{\lambda_1}{ x_i^2} \right ].
\label{bcn}
\eea
\noindent  The spectrum\cite{degu}, 
\be
{\cal{E}}= 2  \omega ( n_1 + \nu_1) + 4 \omega \nu_2,
\ee
\noindent corresponding to the exactly solved states is that of a
superoscillator with the frequency $2 \omega$ and a fermionic
oscillator with the frequency $4 \omega$. The complete spectrum is not
known.

\subsection{Appendix V: Jordan-Wigner transformation}

The spin operators ${S}_i^a, a=1,2,3$ for the $i^{th}$ spin-$\frac{1}{2}$
particle is realized in terms of the Pauli matrices ${\sigma}_i^a$ as,
${S}_i^a :=\frac{1}{2} {\sigma}_i^a$. The Jordan-Wigner transformation
is defined as,
\be
\psi_j = e^{i \pi \sum_{k=1}^{j-1} \sigma_k^+ \sigma_k^- } \sigma_j^-,\ \
\psi_j^{\dagger} = e^{- i \pi \sum_{k=1}^{j-1} \sigma_k^+ \sigma_k^-}
\sigma_j^+ , \ \ \sigma_i^{\pm}:=\frac{1}{2} (\sigma_i^1 \pm i \sigma_i^2),
\label{jw}
\ee
\noindent which relates $\sigma_i^a$'s or the spin operators to the
fermionic variables $\psi_i$'s. The inverse transformation of Eq. (\ref{jw})
is given by,
\be
\sigma_j^- = e^{ i \pi \sum_{k=1}^{j-1} \psi_k^{\dagger} \psi_k} \psi_j, \ \
\sigma_j^+ = e^{ - i \pi \sum_{k=1}^{j-1} \psi_k^{\dagger} \psi_k}
\psi_j^{\dagger}.
\label{ijw}
\ee
\noindent The Jordan-Wigner transformation implies the following identities:
\be
\sigma_i^+ \sigma_{i+1}^-= \psi_i^{\dagger} \psi_{i+1}, \ \
\sigma_i^- \sigma_{i+1}^+ = - \psi_i \psi_{i+1}^{\dagger}, \ \
\psi_i^{\dagger} \psi_i = \sigma_i^+ \sigma_i^- = \frac{1}{2} (1 + \sigma_i^z),
\label{id}
\ee
\noindent which can be used to map a XX spin-chain
system with nearest-neighbour interaction to a system of free fermions.


\end{document}